\documentclass[fleqn,usenatbib,useAMS]{mnras}

\usepackage{newtxtext,newtxmath}
\usepackage[T1]{fontenc}

\DeclareRobustCommand{\VAN}[3]{#2}
\let\VANthebibliography\thebibliography
\def\thebibliography{\DeclareRobustCommand{\VAN}[3]{##3}\VANthebibliography}

\usepackage{booktabs}

\usepackage[dvips]{graphicx}	% Including figure files
\usepackage{amsmath}	% Advanced maths commands

\newcommand{\othreehb}{\rm $\log({}[OIII]\lambda5007/H\beta)$}
\newcommand{\ntwoha}{\rm $\log({}[NII]\lambda6584/H\alpha)$}

\newcommand{\beq}{\begin{equation}}
\newcommand{\eeq}{\end{equation}}
\newcommand{\beqa}{\begin{eqnarray}}
\newcommand{\eeqa}{\end{eqnarray}}

\newcommand{\srg}{{\it SRG}}
\newcommand{\art}{ART-XC}

\newcommand{\swift}{{\it Swift}}
\newcommand{\xrt}{XRT}

\newcommand{\gaia}{{\it Gaia}}
\newcommand{\wise}{{\it WISE}}

\newcommand{\srga}{SRGA\,J230631.0+155633}
\newcommand{\srgas}{SRGA\,J2306+1556}
\newcommand{\sdss}{SDSS\,J230630.38+155620.4}

\def\fmean{\langle f\rangle}

\newcommand{\NH}{N_{\rm H}}
\newcommand{\Lx}{L_{\rm X}}
\newcommand{\Lbol}{L_{\rm bol}}
\newcommand{\Lhx}{L_{\rm HX}}

\newcommand{\Mgal}{M_{\rm gal}}
\newcommand{\tausfh}{\tau_{\rm SFH}}
\newcommand{\tage}{t_{\rm age}}
\newcommand{\Mbh}{M_{\rm BH}}
\newcommand{\Mbulge}{M_{\rm bulge}}
\newcommand{\Msun}{M_\odot}

\title[SRGA\,J2306+1556]{SRGA\,J2306+1556: an extremely X-ray luminous, heavily obscured, radio-loud quasar at $z=0.44$ discovered by \srg/\art}

\author[G.S. Uskov et al.]{
G.S. Uskov,$^{1}$\thanks{Contact e-mail: \href{mailto:uskov@cosmos.ru}{uskov@cosmos.ru}}
S. Sazonov,$^{1}$ I. Lapshov,$^{1}$ A.G. Mikhailov,$^{2}$ E. Filippova,$^{1}$ A. Lutovinov,$^{1}$, I.A. Mereminskiy,$^{1}$   
\newauthor
M. Mochalina,$^{1}$ A.N. Semena,$^{1}$ and A. Tkachenko$^{1}$
\\
$^{1}$Space Research Institute of the Russian Academy of Sciences, Profsoyuznaya Str. 84/32, 117997 Moscow, Russia \\
$^{2}$Special Astrophysical Observatory of the Russian Academy of Sciences, 369167 Nizhnii Arkhyz, Russia \\
}

\date{Accepted XXX. Received YYY; in original form ZZZ}

\pubyear{\the\year{}}

\newcommand{\CellWithForceBreak}[2][c]{
\begin{tabular}[#1]{@{}c@{}}#2\end{tabular}}

\begin{document}
\label{firstpage}
\pagerange{\pageref{firstpage}--\pageref{lastpage}}
\maketitle

% Abstract of the paper
\begin{abstract}
We report on a detailed study of a luminous, heavily obscured ($\NH\sim2\times 10^{23}$\,cm$^{-2}$), radio-loud quasar \srga\, discovered in the 4--12\,keV energy band by the Mikhail Pavlinsky \art\ telescope aboard the \srg\ observatory during the first two years of its all-sky X-ray survey in 2020--2021. The object is located at $z=0.4389$ and is a type 2 AGN according to optical spectroscopy (SDSS, confirmed by DESI). We combine radio-to-X-ray data, including near-simultaneous \art\ and \swift/\xrt\ observations conducted in June 2023. During these follow-up observations, the source was found in a significantly fainter but still very luminous state ($\Lx=1.0^{+0.8}_{-0.3}\times 10^{45}$\,erg\,s$^{-1}$, absorption corrected, 2--10\,keV) compared to its discovery ($\Lx=6^{+6}_{-3}\times10^{45}$\,erg\,s$^{-1}$), which indicates significant intrinsic variability on a rest-frame time scale of $\sim 1$\,year. The radio data show a complex morphology with a core and two extended radio lobes, indicating a giant FRII radio galaxy. From multi-wavelength photometry and the black hole--bulge relation we infer a bolometric luminosity of $\sim 6\times10^{46}$ erg s$^{-1}$ and a black hole mass of $\sim1.4\times10^{9}\,\Msun$, implying accretion at $\sim30$\% of the Eddington limit. \srga\ proves to be one of the most luminous obscured quasars out to $z=0.5$. As such, it can serve as a valuable testbed for in-depth exploration of the physics of such objects, which were much more abundant in the younger Universe.

\end{abstract}

% Select between one and six entries from the list of approved keywords.
% Don't make up new ones.

%%%%%%%%%%%%%%%%%%%%%%%%%%%%%%%%%%%%%%%%%%%%%%%%%%

%%%%%%%%%%%%%%%%% BODY OF PAPER %%%%%%%%%%%%%%%%%%

%%%%%%%%%%%%%%%%%%%%%%
\section{Introduction}
%%%%%%%%%%%%%%%%%%%%%%

Active galactic nuclei (AGN) range greatly in their luminosity, with the most powerful ones (excluding blazars) emitting more than $10^{47}$\,erg\,s$^{-1}$ across the electromagnetic spectrum. Various types of extragalactic surveys indicate that luminous quasars were much more abundant in the past (at $z\gtrsim 1$) than at the present epoch (e.g. \citealt{Ross2013,Ueda2014,Aird2015,Shen2020}). This is usually interpreted in terms of strong evolution of supermassive black hole (SMBH) growth, namely that the most massive holes experienced their major episode(s) of accretion in the first few billion years after the Big Bang (e.g. \citealt{Shankar2009,Hirschmann2014,Weinberger2018}).

A significant fraction of the cosmic SMBH growth has been taking place in obscured AGN, for which the primary (ultraviolet) emission component, originating in the accretion disc, is hidden from our view (mainly due to orientation) by an optically thick layer of cold gas and dust, usually attributed to a so-called torus (see \citealt{Hickox2018} for a review). The torus also absorbs the soft X-rays from the hot corona of the accretion disc. As a result, obscured AGN can only reveal themselves in the infrared (IR) and hard X-ray bands, as well as in (relatively weak) narrow optical emission lines, which complicates their detection and exploration. One of the most important and still open questions is the relative fraction of obscured AGN. There are strong indications, especially in the (better studied) low-redshift Universe, that this fraction decreases with increasing luminosity and Eddington ratio (e.g. \citealt{Hasinger2008,Merloni2014,Ueda2014,Sazonov2015,Ricci2022}).

Due to the combined effect of the steep AGN luminosity function and the decreasing fraction of obscured AGN with luminosity, luminous obscured AGN are extremely rare objects and are thus usually found only at large (cosmological) distances, which complicates their detailed studies. It is thus interesting to search for and explore the nearest of such obscured quasars, for which sufficient photon statistics can be collected by telescopes. In this paper, we describe such a rare case. 

The object of our investigation is a heavily obscured quasar, \srga\ (hereafter, \srgas), which was discovered via blind search during the all-sky X-ray survey \citep{pavlinsky2022} by the Mikhail Pavlinsky \art\ telescope \citep{pavlinsky2021} aboard the \srg\ observatory \citep{sunyaev2021}. We have associated this bright ($\sim 5\times 10^{-12}$\,erg\,s$^{-1}$\,cm$^{-2}$ in the 4--12\,keV energy band) X-ray source with the optical object \sdss, for which an archival spectrum is available from the Sloan Digital Sky Survey (SDSS) DR16 data release \citep{sdss_dr16} that clearly identifies it as a type 2 quasar 
\citep{pavlinsky2022,Sazonov2024} at a fairly low redshift of $0.4386$. This implies that its intrinsic X-ray luminosity exceeded $4\times 10^{45}$\,erg\,s$^{-1}$ during the \srg/\art\ all-sky survey. Furthermore, \srga\ proves to be radio-loud, based on archival radio data. To study this interesting quasar in detail, we organized its follow-up pointed X-ray observations with \srg/\art\ and the \xrt\ telescope \citep{burrows2005} aboard the Neil Gehrels \swift\ observatory \citep{gehrels2004}, which revealed a strongly absorbed X-ray spectrum. 

Below, we examine the physical properties of \srgas\ using the available multi-wavelength information. To compute luminosities, we use a cosmological model with $H_0=70$\,km\,s$^{-1}$\,Mpc$^{-1}$, $\Omega_\Lambda = 0.7$ and $\Omega_{\rm m}=0.3$.

%%%%%%%%%%%%%%%%%%%%%%%%%%%%%%%%%%%%%%%%%%%%%%
\section{X-ray observations and data analysis}
\label{s:xray}
%%%%%%%%%%%%%%%%%%%%%%%%%%%%%%%%%%%%%%%%%%%%%%

\subsection{Discovery and observations during the \srg/\art\ all-sky survey}

\begin{table}
\caption{\label{tab:obs} 
Log of X-ray observations of \srgas.
}
\centering
\begin{tabular}[t]{lcc}
\toprule
Date & Observatory/Telescope/Mode & Exposure, s\\
\midrule
 2020-06-03  & \srg/\art/survey & 23  \\
 2020-12-07  & \srg/\art/survey & 22 \\
 2021-06-08  & \srg/\art/survey & 25 \\
 2021-12-11  & \srg/\art/survey & 24 \\
 2023-06-03   & \srg/\art/pointed & 72900 \\
 2023-06-09   & \swift/XRT/pointed & 2300 \\
 2023-12-13   & \srg/\art/survey & 26 \\
 2024-06-15   & \srg/\art/survey &  26 \\
 2024-12-17   & \srg/\art/survey &  26 \\
\bottomrule
\end{tabular}
\label{tab:xray-obs}
\end{table}

The X-ray source was discovered in the 4--12\,keV energy band by the \art\ telescope during the first year of the \srg\ all-sky survey, on the summed map of the first two half-year scans (ARTSS1-2, \citealt{pavlinsky2022}). Its existence was then confirmed in the updated source catalogue based on the data of the first four \art\ scans (ARTSS1-5, \citealt{Sazonov2024})\footnote{This catalogue also includes data of the fifth scan for $\sim 40$\% of the sky, hence the name ARTSS1-5.}. The source is listed as SRGA\,J230630.9+155637 and SRGA\,J230631.0+155633 in ARTSS1-2 and ARTSS1-5, respectively. To our knowledge, it is not present in any published X-ray source catalogues from other missions.  
The statistical positional uncertainty of the source in ARTSS1-5 is $R_{98}=23.2\arcsec$ (the radius of the 98\% confidence region), as shown in Fig~\ref{fig:img-xray}.

In October 2023, \art\ resumed the all-sky survey, and as of writing, \srgas\ has been scanned three more times since then. Table~\ref{tab:obs} provides the dates and vignetting corrected exposures of all seven visits of \srgas\ during the \srg/\art\ all-sky survey.

\subsection{\srg/\art\ pointed observation}

To obtain X-ray spectral information, pointed observations of \srgas\ were performed in June 2023, first by \srg/\art\ in hard X-rays and six days later by \swift/\xrt\ in soft X-rays. See Table \ref{tab:obs} for more details. 

During the \art\ observation (ID 12310089001, PI: S. Sazonov), the telescope was pointed at the presumed optical counterpart \sdss. The telescope's field of view (FoV) is 36 arcminutes in diameter, which provides a comprehensive picture of the surrounding region. All seven detector modules were operational during the observation. A few bad time intervals, amounting in total to nearly 300 seconds, were excised from the obtained data.

Using attitude data from the on-board inertial navigation system GYRO, we located the X-ray source at RA=23h06m30.4s, DEC=+15\degr 56\arcmin 17\arcsec, with a 98\% statistical error $r_{98}=2\arcsec$. We determined this confidence region based on likelihood fitting with the point-spread function (PSF). Given the current quality of the \art\ PSF model, the likelihood was computed for sky maps with a 2\arcsec\ pixel size and the 98\% confidence interval fell within the pixel. In addition, a 7\arcsec\ systematic uncertainty caused by the attitude precision must be included. Thus, the total position uncertainty of \srgas\ can be estimated at 7.3\arcsec.

\begin{figure*}
    \centering
    \includegraphics[width=0.68\columnwidth]{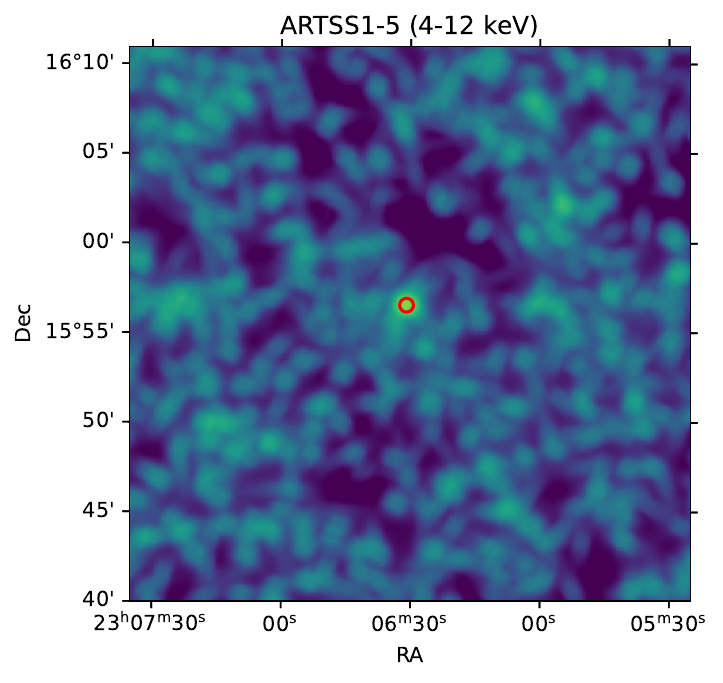}
    \includegraphics[width=0.68\columnwidth]{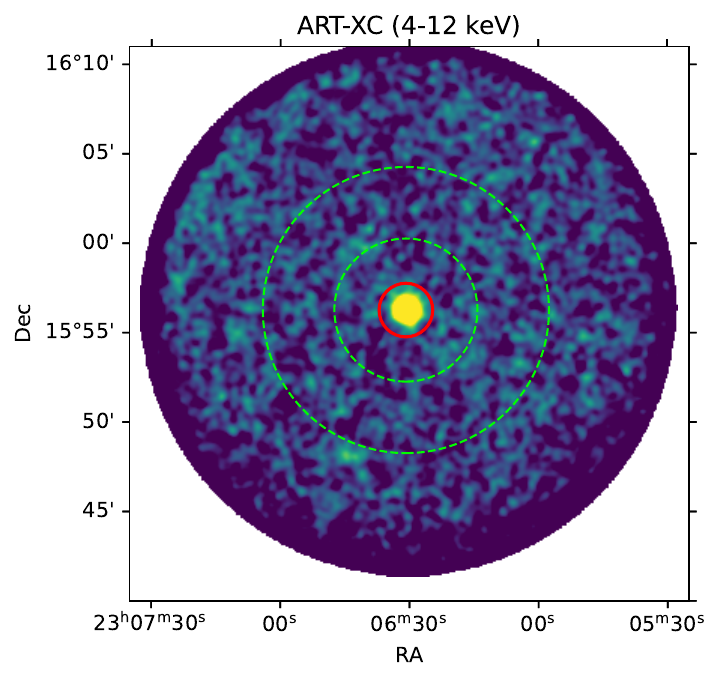}
    \includegraphics[width=0.68\columnwidth]{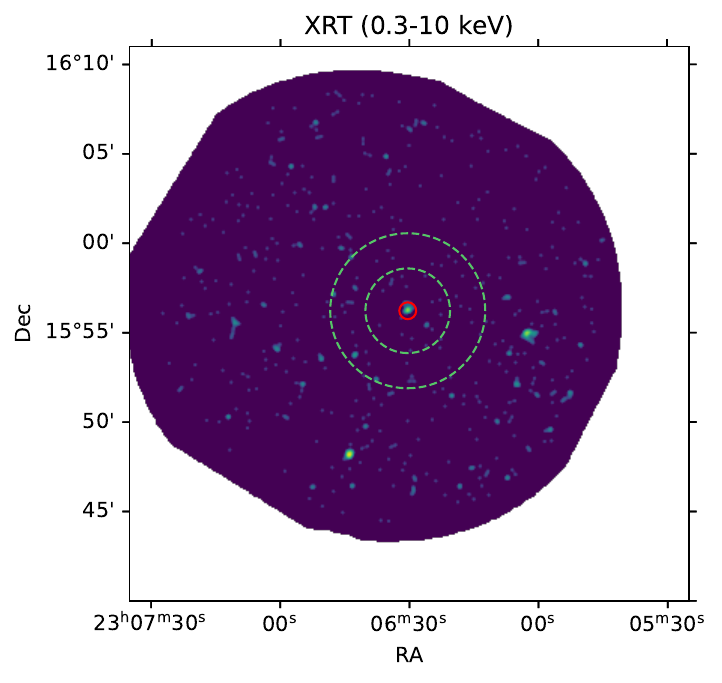}
    \caption{    
    Smoothed X-ray images of \srgas. The left image is based on survey data from ARTSS1-5 (4--12\,keV), while the middle and right images are based on pointed observations by \srg/\art\ and \swift/\xrt\ (see Table~\ref{tab:obs}) in the 4--12 and 0.3--10\,keV bands, respectively. In the left panel, the red circle indicates the 98\% localization region, with a radius of 23.2\arcsec; the circle is centered on the ARTSS1-5 coordinates. In the middle and right panels, the red solid circle marks the source spectrum extraction region, with a radius of 90\arcsec\ for ART-XC and 28.32\arcsec\ for \xrt, respectively. The green dashed annuli represent the background extraction region, with an inner radius of 4\arcmin\ and outer radius of 8\arcmin\ for ART-XC, and 142\arcsec\ and 260\arcsec\ for XRT, respectively. The \art\ source region is centered on the optical counterpart \sdss, while the \xrt\ source region is centered on the X-ray source (these positions are different by just 6.5 arcseconds). 
    }
    \label{fig:img-xray}
\end{figure*}

We extracted the source and background spectra using the standard \art\ software package, adapted to the goals of this study with the current version of the calibrations. We constructed the photon, exposure, and vignetting maps for individual energy intervals. The spectrum of \srgas\ was extracted from a circle of 90\arcsec\ radius centered on the quasar's optical coordinates. The background spectrum was estimated within an annular region with an inner and outer radius of 4\arcmin\ and 8\arcmin, respectively (Fig.~\ref{fig:img-xray}). The \art\ background is strongly dominated by the particle contribution, with the cosmic (extragalactic) X-ray background contributing just $\sim 3$\% in the 4--12\,keV band and even less at higher energies \citep{Sazonov2024}. The particle background is actually weakly dependent on the position within the \art\ FoV. Using background model maps obtained during the all-sky survey in `blank' fields, we verified that the assumption of a flat background does not significantly affect the inferred source photon counts for \srgas\ across the spectrum. We thus ignore this effect and use the background spectrum in our spectral analysis in the standard way, i.e. just taking into account the difference in the areas of the background and source regions.

At 9\arcmin\ from \srgas, \art\ detected another source  (23h06m44s, +15d48m11s) during the pointed observation. 

Its observed flux in the 4--12\,keV energy band is $\sim2\times10^{-13}$\,erg\,s$^{-1}$\,cm$^{-2}$ and the detection likelihood is 41.2. Due to its faintness and large offset, this source does not affect our estimation of the source and background spectra for \srgas. 

We limited our spectral analysis to the 4--20\,keV energy band, since the background dominates over the source signal above 20\,keV.

To study the short-term variability of \srgas\ during the pointed observation, we used several energy bands between 4 and 16\,keV. To this end, the source counts were extracted from a circular region with a radius of 1\arcmin\ centered on \sdss. 
We did not subtract background counts for this analysis, since the \art\ background is very stable on the relevant time scales.

\subsection{\swift/\xrt\ pointed observation}

The source was observed by the \swift\ observatory (Target of Opportunity observation ID 18931, requested by I.~Mereminskiy). We processed the \swift/\xrt\ data (Target ID 00016070) using the University of Leicester's service\footnote{www.swift.ac.uk/user\_objects}, which utilizes HEASOFT v6.32\footnote{www.heasarc.gsfc.nasa.gov/docs/software/heasoft}. 

The telescope was pointed at \sdss. The target source is confidently detected in the \swift/\xrt\ image (see Fig.~\ref{fig:img-xray}), with a position RA=23h06m30.4s, DEC=$15\degr 56\arcmin 14\arcsec$. The standard position error provided by the \swift\ pipeline software based on the star tracker is 6.3\arcsec, which is composed of the 90\% statistical uncertainty, $r_{90}=5.2\arcsec$, and the systematic error of 3.5\arcsec\ added in quadrature. Unfortunately, it is not possible to derive an `enhanced position' for the source, because \swift\ was in an anomaly state during the observation of \srgas\  (Leicester Swift Helpdesk, private communication) due to problems with the attitude control system, which affected the data of the UVOT telescope \citep{roming2005}. Assuming a Gaussian profile for the source, we can estimate the statistical position uncertainty at the 98\% confidence level (as for \art\ above) at $r_{98}\approx 1.3 r_{90}=6.8\arcsec$, and adding the aforementioned systematic error we obtain a conservative estimate of the total uncertainty in the source position of 7.6\arcsec. 

To extract spectra, we used grades 0--12 in Photon Counting (PC) mode. We set the centroid method to `single pass' with a maximum of 10 attempts. The source spectrum was extracted from a circle of $28.3\arcsec$ radius centered on the position of the X-ray source. The background spectrum was extracted from an annulus between 142\arcsec\ and  260\arcsec\ from the source (Fig.~\ref{fig:img-xray}). We restricted our spectral analysis to the 0.3--10\,keV energy band, as recommended for \xrt. 

There are no other sources within the source and background extraction regions. The nearest source detectable by the standard pipeline is located at 6\arcmin\ from \srgas, which ensures that no external contamination affects the extracted spectra.

%%%%% 
\subsection{X-ray spectrum}
\label{s:xrayspec}

We performed X-ray spectral modeling using \emph{XSPEC} (version 12.12.0, \citealt{arnaud96}). The \swift/\xrt\ and \srg/\art\ were analyzed jointly, using C-statistic 
%(as dictated by the \art\ background estimation procedure) 
to evaluate the quality of spectral modeling. 
We grouped the \xrt\ data to ensure a minimum of three counts per energy bin in the source spectrum. The resulting \xrt\ source spectrum covers the 0.3--5.73\,keV energy band and contains a total of nine counts.
The \art\ data were grouped with a minimum of 20 counts per bin in the source spectrum. There are a total of 2701 counts in the \art\ source spectrum over the 4--20\,keV energy band.

\begin{figure}
    \centering
    \includegraphics[width=1\columnwidth]{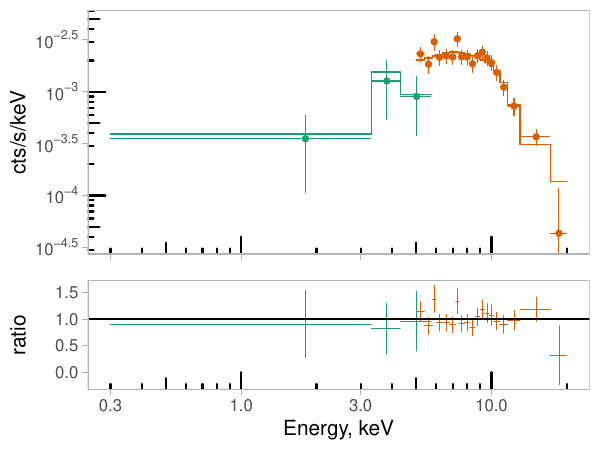}
    \caption{
    The upper panel shows the X-ray spectrum of \srgas\ measured with \srg/\art\ (orange) and \swift/\xrt\ (green), and fitted by the absorbed power-law model. The \art\ spectrum is re-binned in the plot for better clarity. The bottom panel shows the data-to-model ratio. The model parameters are listed in Table~\ref{fig:art_swift_spec}. The uncertainties are shown at $1\sigma$ confidence level.
    }
    \label{fig:art_swift_spec}
\end{figure}

According to the widely accepted AGN paradigm, Comptonization of the accretion disc's photons in its hot corona leads to the formation of a power-law spectrum with a high-energy cutoff \citep{sunyaev80,haardt91}, typically above $\sim 100$\,keV (e.g. \citealt{Malizia2020}). Also a low-energy cutoff appears in the X-ray spectrum due to photoelectric absorption. Since the spectral range of the \art\ data is limited in our analysis by a maximum energy of 20\,keV ($\approx29$~keV in the \srgas\ rest frame), we do not consider a high-energy cutoff and use a simple power-law model modified by absorption in the Galaxy and within the quasar. Hence, we use the following spectral model (in the \emph{XSPEC} terminology):
\begin{equation}
\text{\textsc{tbabs}} \times \text{\textsc{ztbabs}} \times \text{\textsc{cflux}} \times \text{\textsc{zpowerlaw}}.
\label{eq:pl}
\end{equation}
Here, {\sc tbabs} is the Galactic interstellar absorption model \citep{wilms2000}. We adopt the Galactic hydrogen column density in the direction of \srgas\ from the HI4PI survey data \citep{bekhti16}, $N_{\rm H}^{MW}=5.7\times 10^{20}$\,cm$^{-2}$, and the solar element abundances. To describe the intrinsic absorption in the object ($N_{\rm H}$), we use the {\sc ztbabs} model, again assuming the standard chemical composition. The power-law continuum is characterized by the photon index $\Gamma$ and the absorption-corrected flux in the observed 4--10\,keV band ({\sc cflux}) $F_{\rm 4-10}^{PL}$. 

The \xrt\ and \art\ joint spectrum can be described by this model quite well, as shown in Fig.~\ref{fig:art_swift_spec}. The derived spectral parameters are presented in Table~\ref{tab:xray-params}. 

\begin{table*}
\caption{
\label{tab:xray-params} 
X-ray spectral parameters.
}
\centering\centering

\begin{tabular}{lllllllll}
\toprule
$N_{\rm H}^{MW}$ & $N_{\rm H}$ & $F_{\rm 4-10}^{PL}$ & $\Gamma$ & $F_{\rm 4-10}^{PEXRAV}$ & $E$ & $EW$ & $\sigma_{\rm line}$ & $Cstat$ (dof)\\
\addlinespace[0.5em]
$10^{22}$\,cm$^{-2}$ & $10^{22}$\,cm$^{-2}$ & $10^{-13}$\,erg\,s$^{-1}$\,cm$^{-2}$ & & $10^{-13}$\,erg\,s$^{-1}$\,cm$^{-2}$ & keV & keV & keV & \\
\midrule
\multicolumn{9}{l}{\sc \textbf{tbabs$\times$ztbabs$\times$cflux$\times$zpowerlaw}}\\
\hspace{1em}$0.057$ (fixed) & $24_{-15}^{+31}$ & $8_{-2}^{+5}$ & $2.0_{-0.5}^{+0.7}$ &  &  &  &  & 57.4 (65)\\
\addlinespace[0.3em]
\multicolumn{9}{l}{\sc \textbf{tbabs(ztbabs$\times$cflux$\times$zpowerlaw + cflux2$\times$pexrav)}}\\
\hspace{1em}$0.057$ (fixed) & $18_{-8}^{+10}$ & $7.6_{-1.5}^{+0.7}$ & $1.80$ (fixed) & $< 0.6$ &  &  &  & 57.8 (65)\\
\addlinespace[0.3em]
\multicolumn{9}{l}{\sc \textbf{tbabs(ztbabs$\times$cflux$\times$ zpowerlaw + zgauss)}}\\
\hspace{1em}$0.057$ (fixed) & $19_{-8}^{+11}$ & $7.6_{-0.8}^{+0.9}$ & $1.80$ (fixed) &  & $6.4$ (fixed) & $< 1.3$ & $0.01$ (fixed) & 57.4 (65)\\
\bottomrule
\end{tabular}
\begin{flushleft}
The uncertainties and upper limits for the parameters are quoted at the 90\% level of confidence.

\end{flushleft}
\end{table*}

\begin{figure}
    \centering
    \includegraphics[width=1\columnwidth]{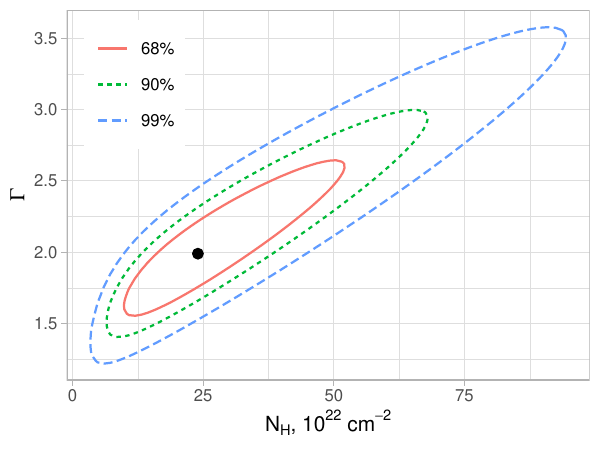}
    \caption{
    Joint confidence regions for the $N_{\rm H}$ and $\Gamma$ parameters of the {\sc tbabs$\times$ztbabs$\times$cflux$\times$zpowerlaw} model at the 68\%, 90\% and 99\% levels.
    }
    \label{fig:pow_nh_gamma}
\end{figure}

We confidently detect intrinsic absorption in the object, although the absorption column density is poorly constrained by the data because of the degeneracy with the photon index (see Fig.~\ref{fig:pow_nh_gamma}). Specifically, $N_{\rm H}$ can be anywhere between $10^{23}$\,cm$^{-2}$ and $6\times 10^{23}$\,cm$^{-2}$ (90\% confidence). 

The inferred slope of the power-law continuum, $\Gamma=2.0_{-0.5}^{+0.7}$, is consistent with typical values found for AGN. Given the presence of the intrinsic absorption (presumably, due to a dusty torus), it is natural to expect also the presence of a Compton reflection component. Therefore, we added a {\sc pexrav} \citep{Magdziarz1995} reflection component to our model, i.e.:
\begin{equation}
\text{\textsc{tbabs}}\,(\text{\textsc{ztbabs}} \times \text{\textsc{cflux}} \times \text{\textsc{zpowerlaw}} + \text{\textsc{cflux2}} \times \text{\textsc{pexrav}})
\label{eq:pl_pex}
\end{equation}

We fixed the photon index at 1.8 for both the {\sc zpowerlaw} and {\sc pexrav} components. To include only the reflected emission (without the direct emission continuum) in {\sc pexrav}, we fixed the rel\_refl parameter at a negative value. As before, the model does not include a high-energy cutoff. The cosine of the inclination angle was fixed at 0.5. The abundances of all elements heavier than He were fixed at 1.0 relative to the solar ones, adopted from \citet{anders&grevesse1989}. The only free parameter of the reflection component is its flux in the observed 4--10\,keV band corrected for Galactic absorption ({\sc cflux2}), $F_{\rm 4-10}^{PEXRAV}$.

As shown in Table~\ref{tab:xray-params}, reflection emission is not detected in the spectrum, and it adds less than 8\% (90\% confidence) to the intrinsic (i.e. corrected for intrinsic absorption) power-law continuum in the observed 4--10\,keV energy band (5.8--14.4\,keV in the quasar's rest frame) or, equivalently, contributes less than 11\% of the total observed emission in this band. For the adopted spectral slope $\Gamma=1.8$, the absorption column is constrained fairly tightly between $1.0\times 10^{23}$ and $2.8\times 10^{23}$\,cm$^{-2}$. Within the AGN torus paradigm, one expects the reflected component to contribute less than 15\% (the exact number depends on the line-of-sight and equatorial column densities as well as other parameters of the torus) to the total observed emission in the 5.8--14.4\,keV energy band for such moderate (Compton thin) column densities (e.g. \citealt{Melazzini2023}), which is consistent with the upper limit inferred for \srgas.

In the same scenario of reflection of X-rays from the torus, there must also arise a K$\alpha$ fluorescent line of iron at $\sim6.4$\,keV. We thus added a Gaussian component {\sc zgauss} at the rest-frame energy $E=6.4$\,keV with a free normalization %$A^{GAUSS}$ 
or, equivalently, with a free equivalent width, $EW$:
\begin{equation}
\text{\textsc{tbabs}} \left( \text{\textsc{ztbabs}} \times \text{\textsc{cflux}} \times \text{\textsc{zpowerlaw}} + \text{\textsc{zgauss}} \right).
\label{eq:pl_gauss}
\end{equation}
The photon index was fixed as before at $\Gamma=1.8$. 

The fit proves to be nearly insensitive to the width of the Gaussian for $\sigma_{\rm line} <0.5$\,keV.
We thus fixed it at $0.01$\,keV to present the results in Table~\ref{tab:xray-params}. The line is not detected, and the 90\% upper limit on its equivalent width is 1.3\,keV. This is consistent with the expected strength of the 6.4\,keV line $EW<200$\,eV for tori with $N_{\rm H}\lesssim 3\times 10^{23}$\,cm$^{-2}$ and solar abundance of iron \citep{Melazzini2023}.

\subsection{Short-term X-ray variability}

\begin{figure}
    \centering
    \includegraphics[width=1\columnwidth]{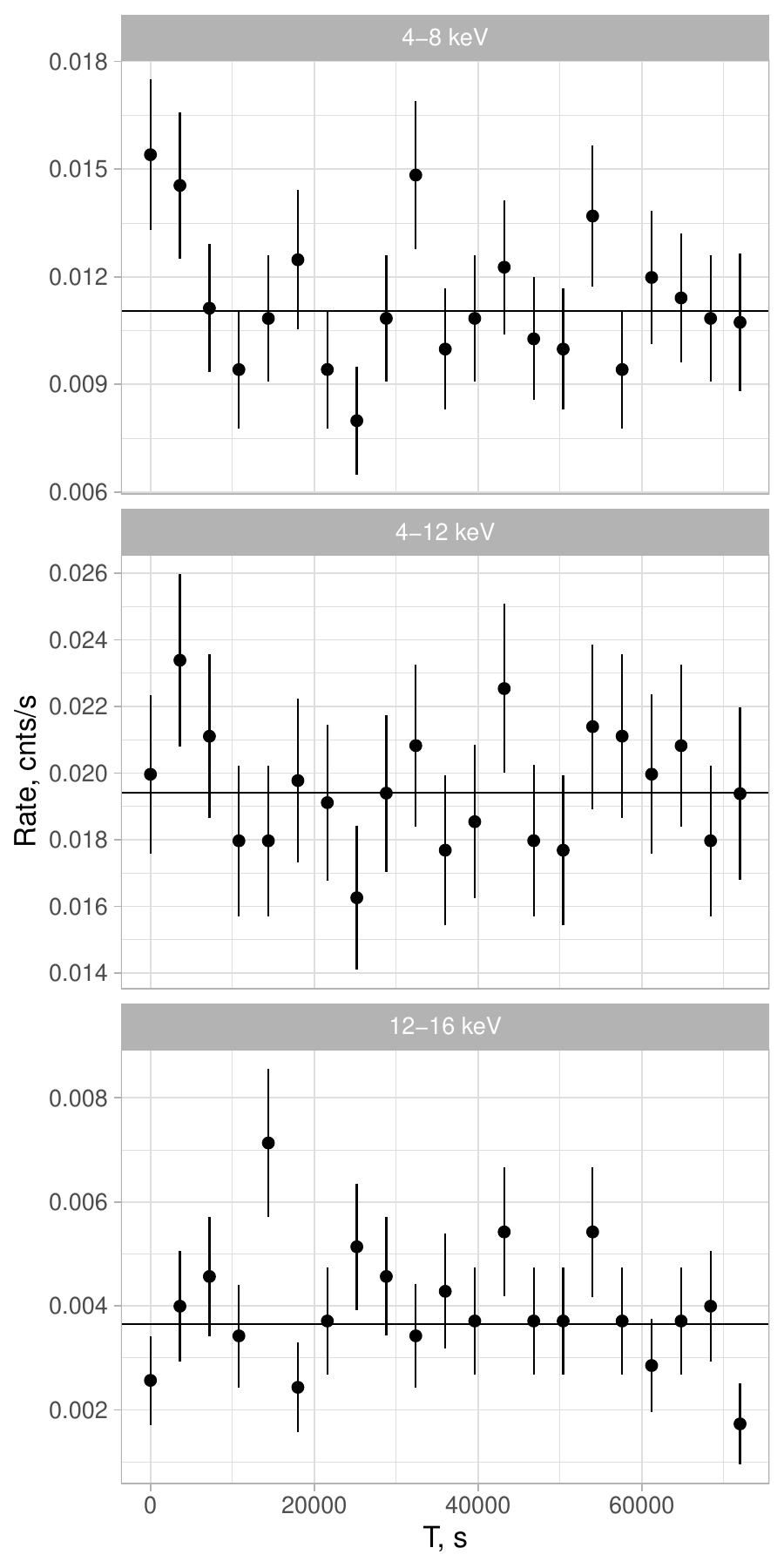}
    \caption{X-ray light curves of \srgas\ in three energy bands obtained by the \art\ telescope during the pointing observation, with a bin time of one hour. The error bars represent the 1$\sigma$ uncertainty level. 
    }
    \label{fig:artxc_lc}
\end{figure}

We next examine the short-term (intraday) X-ray variability of \srgas. To this end, we constructed light curves (count rate as a function of time, in 1-hour bins) during the \art\ pointed observation in the main energy band of 4--12\,keV and in two additional bands: 4--8 and 12--16\,keV (Fig.~\ref{fig:artxc_lc}). Because \swift/\xrt\ detected just ten counts in the 0.3--10\,keV energy band from the source region during its short ($\approx 40$~min) observation, we did not use these data for studying variability.

To investigate whether the X-ray flux is variable in a given energy band during the observation, we calculate a $\chi^2$ statistic\footnote{The light curves contain $\gtrsim$ 25 counts per bin in the 4--8\,keV and 4--12\,keV bands and $\gtrsim$10 counts per bin in the 12--16\,keV band, so that the Gaussian approximation is adequate for this analysis.}, as follows:
\begin{equation}
    \chi^2 = \sum_{i=1}^{N} \frac{(f_{\rm i} - \overline{f})^2}{\sigma_{\rm i}^2},
\end{equation}
where $f_{\rm i}$ is the count rate in the {\it i}-th time bin, $\sigma_{\rm i}$ is the corresponding error, $\overline{f}$ is the weighted mean count rate for the observation, and $N$ is the total number of measurements. Accordingly, we determine $1-p$ -- the probability of obtaining by chance a $\chi^2$ value greater than the measured one for ${\rm dof}=N-1$ degrees of freedom under the hypothesis that the count rate remained constant during the observation. 

To estimate the amplitude of X-ray variability, we calculate the normalized excess variance, which is the variance normalized to the mean flux and corrected for the flux measurement errors \citep{vaughan2003}:
\begin{equation}
    \sigma^2_{\rm rms} = \frac{S^2 - \overline{\sigma^2_{\rm err}}}{\fmean^2},
\end{equation}
where
\begin{equation}
    S^2 = \frac{1}{N-1}\sum_{i=1}^{N}(f_{\rm i} - \fmean)^2
\end{equation}
and
\begin{equation}
    \overline{\sigma^2_{\rm err}} = \frac{1}{N}\sum_{i}^N \sigma_{\rm i}^2.
\end{equation}
The error of $\sigma_{\rm rms}^2$ can be estimated as follows \citep{turner1999}:
\begin{equation}
\delta\sigma_{\rm rms}^2=\frac{s_D}{\fmean^2\sqrt{N}},
\end{equation}
where
\begin{equation}
s^2_D = \frac{1}{N-1} \sum_{i=1}^{N}\{[(f_{\rm i} - \fmean)^2 - \sigma^2_{\rm i}] - \sigma^2_{\rm rms}\fmean^2\}^2.
\label{eq:sigmaerr}
\end{equation}

Table~\ref{tab:time_stat} presents the derived mean flux rate $\overline{R}$, $\chi^2 ({\rm dof})$, $1-p$, and $\sigma_{\rm rms}^2$ for each energy band. The X-ray flux did not exhibit any statistically significant variations (i.e., $1-p>0.05$) over the $\sim 20$-hour period covered by the \art\ observation. 

\begin{table}
\caption{
\label{tab:time_stat} 
Characteristics of the short-term X-ray variability.
}
\begin{tabular}{lllll}
\toprule
Band, keV & $\overline{R}$, cnt\,s$^{-1}$ & $\chi^2$ (dof) & $1-p$ & $\sigma^2_{\rm RMS}$\\
\midrule
4--8 & $0.011$ & $21.88$ (20) & $0.35$ & $0.004\pm0.008$\\
4--12 & $0.019$ & $11.39$ (20) & $0.94$ & $-0.0063\pm0.0022$\\
12--16 & $0.004$ & $23.98$ (20) & $0.24$ & $0.02\pm0.03$\\
\bottomrule
\end{tabular}
\end{table}

%%%%% 
\subsection{Long-term X-ray variability}

Using the available data of the all-sky survey and pointed \art\ observations (Table ~\ref{tab:xray-obs}), we can also study the X-ray variability of \srgas\ on time scales of years. To this end, for the survey observations, we performed forced photometry in the circular region with a radius of 71\arcsec\ around the \sdss\ position. We then converted the measured count rates to energy fluxes in the 4--12\,keV energy band assuming a Crab-like spectrum\footnote{As discussed by \cite{Sazonov2024}, the energy flux to count rate conversion factor only weakly depends on the spectral shape.} and applying an aperture correction of 96\%. The 90\% upper limits were calculated using equation~(7) from \citet{gehrels1986}, with $S=1.282$. The corresponding background was estimated using \art\ measurements in the 30--70\,keV energy band (see \citealt{Sazonov2024} for details). The inferred fluxes are given in Table~\ref{tab:time_long_cnts}. 

For the \art\ pointing observation, we estimated the observed flux in the 4--12\,keV energy band using our baseline spectral model {\sc cflux$\times$tbabs$\times$ztbabs$\times$zpowerlaw} (see Table~\ref{tab:xray-params}). It turns out to be $(7.6\pm0.3)\times10^{-13}$\,erg\,s$^{-1}$\,cm$^{-2}$ at the 1$\sigma$ confidence level (see also the last row of Table~\ref{tab:time_long_cnts}).

\begin{figure}
    \centering
    \includegraphics[width=1\columnwidth]{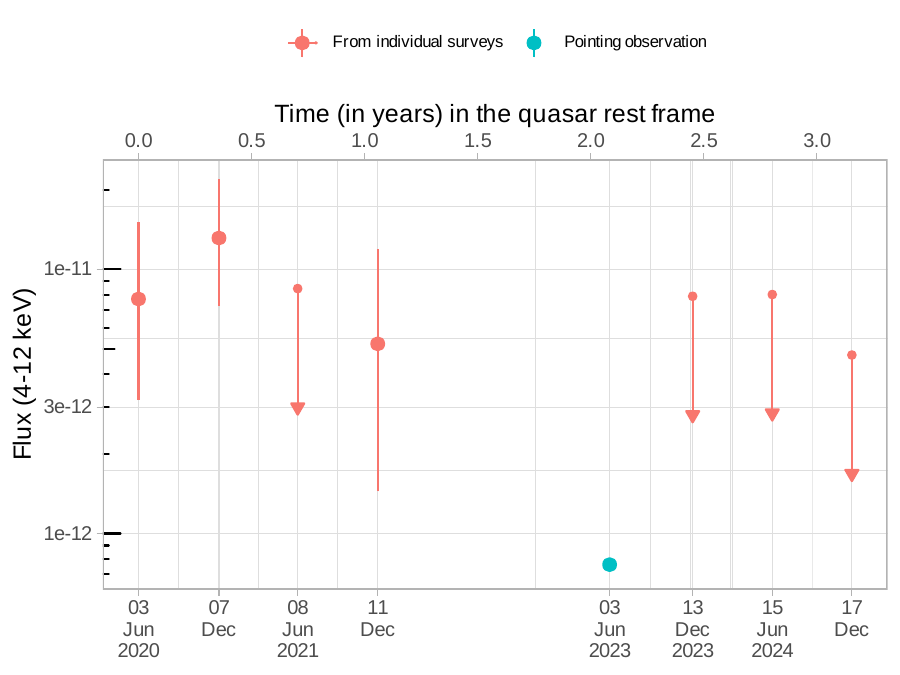}
    \caption{
    Long-term X-ray light curve of \srgas. Shown is the observed flux in the 4--12\,keV energy band. The lower horizontal axis provides the dates of observations, while the upper axis shows the time passed since the first observation of the source in its rest frame. The error bars represent 1$\sigma$ uncertainties, while the upper limits are shown at the 90\% confidence level.
    }
    \label{fig:artxc_longterm_lc}
\end{figure}

\begin{table}
\caption{
\label{tab:time_long_cnts} 
Information on the detection of \srgas\ in the individual \art\ all-sky surveys.
}
\begin{tabular}{ccccc}
\toprule
\art\ survey & $S$ & $B$ & $1-p$ & \CellWithForceBreak{$F_{4-12}$, \\ $10^{-12}$\,erg\,s$^{-1}$\,cm$^{-2}$ }\\
\midrule
1 & 4 & 0.69 & 0.005361 & $8_{-5}^{+7}$\\
2 & 6 & 0.65 & 0.000060 & $13_{-6}^{+9}$\\
3 & 1 & 0.79 & 0.544614 & $<8$\\
4 & 3 & 0.71 & 0.035158 & $5_{-4}^{+7}$\\
6 & 1 & 0.44 & 0.359156 & $<8$\\
7 & 1 & 0.49 & 0.384645 & $<8$\\
8 & 0 & 0.43 & 1.000000 & $<5$\\
\midrule
pointing & & & & $0.76\pm0.03$ \\
\bottomrule
\bottomrule
\end{tabular} 
\begin{flushleft}
$S$ -- number of observed counts; $B$ -- predicted background counts in the source region in the 4--12\,keV energy band; ($1-p$) -- probability of the null hypothesis (i.e. there is no source), assuming Poisson photon statistics: $Pr(N \geqslant S | \lambda=B$); $F_{4-12}$ -- observed flux in the 4--12\,keV energy band. Uncertainties are at the 1$\sigma$ level, and upper limits at 90\% confidence. Note that \srgas\ was not scanned by \art\ during the incomplete fifth sky survey in Dec. 2021--March 2022.
\end{flushleft}
\end{table}

The resulting light curve in the 4--12\,keV energy band is shown in Fig.~\ref{fig:artxc_longterm_lc}. We clearly see strong variability. The source was relatively bright during the \art\ all-sky survey visits between June 3, 2020 and December 11, 2021. The maximum flux $13_{-6}^{+9}~\times10^{-12}$\,erg\,s$^{-1}$\,cm$^{-2}$ was reached in the second survey (on Dec. 7, 2020), although the variability between the first four surveys is statistically insignificant. Formally, the source was detected with high statistical significance in the second survey and less significantly in the first and fourth surveys (see Table~\ref{tab:time_long_cnts}). The detection of \srgas\ on the summed map of the first four \art\ all-sky surveys is also robust (maximum likelihood detection significance $S/N=5.65$) and the corresponding average flux is $(4.6\pm 1.6)\times10^{-12}$\,erg\,s$^{-1}$\,cm$^{-2}$ (4--12\,keV, \citealt{Sazonov2024}). Subsequently, during the pointed observation on June 3, 2024, the quasar was dimmer by a factor of 9--30 compared to the peak flux measured on Dec. 7, 2020. Afterwards, during the resumed all-sky survey, on Dec. 13, 2023, June 15, 2024, and Dec. 17, 2024, the source was not detected anymore, but the derived upper limits on its flux are not very restrictive. 

We can thus conclude that \srgas\ experienced an X-ray outburst in 2020--2021 and was in a `low' state in June 2023. Given the absence of data before 2020 and the gaps in monitoring afterwards, the outburst could have lasted $\sim 1$\,year or more in the quasar's rest frame. Importantly, this variability cannot be explained by variable obscuration. Indeed, according to our X-ray spectral analysis in Section~\ref{s:xrayspec}, intrinsic absorption, although substantial, suppressed the observed 4--12\,keV flux by only a factor of $1.3$ when the source was in its low X-ray state, much smaller than the observed drop in the flux relative to the `high' state caught during the \srg/\art\ all-sky survey. Therefore, the outburst was definitely caused by an increase in the intrinsic luminosity rather than by the disappearance of intervening cold gas. 

It is also worth noting that such an X-ray outburst is expected to be followed by an `echo' due to scattering in the surrounding torus. Although the result will strongly depend on the orientation and structure of the torus, the minimum duration of the reflected signal will be determined by the inner radius of the torus, $R_{\rm dust}$. For an estimated bolometric luminosity of $\Lbol\sim 6\times 10^{46}$\,erg\,s$^{-1}$ (see Sect.~\ref{s:sed} below) of \srgas, one expects $R_{\rm dust}\sim 2$\,pc based on dust sublimation physics and reverberation mapping studies of AGN (e.g. \citealt{Mandal2024}). This corresponds to times scales longer than $\sim 6$\,years. Therefore, the non-detection of a strong reflection feature in the X-ray spectrum of \srgas\ in its low state does not seem surprising.

%%%%%%%%%%%%%%%%%%%%%%%%%%%%%%
\section{Multiwavelength data}
\label{s:multi}
%%%%%%%%%%%%%%%%%%%%%%%%%%%%%%

\begin{figure}
    \centering
    \includegraphics[width=1\columnwidth]{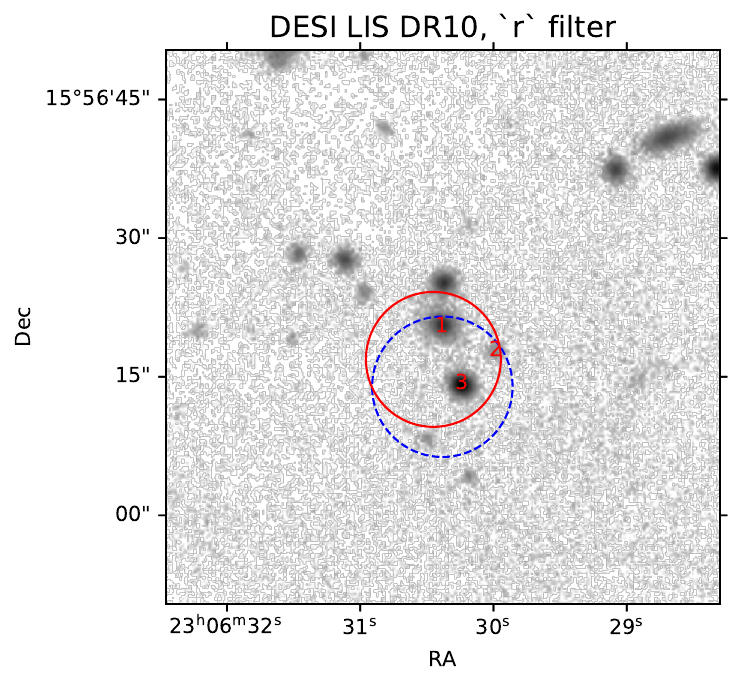}
    \caption{
    Optical \emph{r}-band image from LS DR10. The red solid and blue dashed circles denote the 98\% localization regions of the X-ray source based on the \srg/\art\ and \swift/\xrt\ pointed observations, with radii of 7.3\arcsec and 7.6\arcsec, respectively. The numbers mark optical sources that are present in the overlapping area of these regions (see Table.~\ref{tab:objects}). 
    }
    
\label{fig:desilis}
\end{figure}

\begin{table*}
\caption{
Optical/IR objects within the \srg/\art--\swift/\xrt\ X-ray localization region. 
\label{tab:objects}
}
\setlength{\tabcolsep}{1.5pt}

\setlength{\tabcolsep}{1pt}
\begin{tabular}{cccccccccccccc}
\toprule
\multicolumn{7}{c}{LS} & 
\multicolumn{3}{c}{UKIRT HS} &  
\multicolumn{4}{c}{LS--WISE}  \\
\cmidrule(lr){1-7} \cmidrule(lr){8-10} \cmidrule(lr){11-14}
\multicolumn{1}{c}{Offset} & 
\multicolumn{1}{c}{RA} & 
\multicolumn{1}{c}{DEC} & 
\multicolumn{1}{c}{\emph{g}} & 
\multicolumn{1}{c}{\emph{r}} & 
\multicolumn{1}{c}{\emph{z}} & 
\multicolumn{1}{c}{Type} & 
\multicolumn{1}{c}{\emph{J}} & 
\multicolumn{1}{c}{\emph{K}} & 
\multicolumn{1}{c}{Class} & 
\multicolumn{1}{c}{\emph{W1}} & 
\multicolumn{1}{c}{\emph{W2}} & 
\multicolumn{1}{c}{\emph{W3}} & 
\multicolumn{1}{c}{\emph{W4}} \\
\midrule\addlinespace[2.5pt]
\multicolumn{13}{c}{Object 1} \\
0.0 & 346.62658 & 15.93905 & $21.17 \pm 0.02$ & $19.861 \pm 0.009$ & $18.791 \pm 0.015$ & SER & $17.8\pm 0.2$ & $15.77\pm0.12$ & Gal. & $14.606 \pm 0.008$  & $13.309 \pm 0.009$ & $9.28 \pm 0.02$ & $5.78\pm0.06$ \\
\multicolumn{13}{c}{Object 2} \\
6.3 & 346.62494 & 15.93828 & $24.14 \pm 0.19$ & $22.28 \pm 0.05$ & $21.37 \pm 0.10$ & REX & $-$ & $-$ & $-$ &  $18.09 \pm 0.13$ & $> 18.1$ & $>12.8$ & $ > 8.2$\\
\multicolumn{13}{c}{Object 3} \\
6.7 & 346.62603 & 15.93728 & $21.802 \pm 0.018$ & $19.900 \pm 0.005$ & $17.777 \pm 0.003$ & PSF & $16.19 \pm 0.03$  & $15.26\pm 0.04$ & Gal. &  $16.89 \pm 0.05$ & $15.76 \pm 0.06$ & $13.1 \pm 0.8$ & $ > 8.2$ \\
7.3 & 346.62578 & 15.93717 & $23.84 \pm 0.11$ & $21.95 \pm 0.03$ & $19.727 \pm 0.018$ & PSF &  & & &  $15.328 \pm 0.012$ & $15.67 \pm 0.06$ & $>12.8$ & $10\pm3$\\
\bottomrule
\end{tabular}

\begin{flushleft}
The first seven columns provide information from LS DS10, namely: angular distance (in arcsec) from \sdss, ICRS coordinates, g, r, and z magnitudes (in the AB system), and morphological classification. The next three columns provide the J, K magnitudes (in the Vega system) and classification (Gal. for galaxy) from UHS DR2. The final four columns provide the W1, W2, W3 and W4 magnitudes (in the Vega system) obtained by \wise\ forced photometry at the LS DR10 positions. 
\end{flushleft}
\end{table*}

Figure \ref{fig:desilis} presents a $1\arcmin\times 1\arcmin$ optical image around \srgas\ obtained in the \emph{r} band by the DESI Legacy Surveys (LS, \citealt{dey2019}). There are three optical/IR objects within the overlapping area of the \srg/\art\ and \swift/\xrt\ localizations regions. Table~\ref{tab:objects} presents photometric data from the LS Data Release 10 (DR10) and the UKIRT Hemisphere Survey Data Release 2 (UHS DR2, \citealt{dye2018}). 

Apart from optical photometry (\emph{g},\emph{r},\emph{z}), LS DR10 provides mid-infrared (MIR) photometry from the {\it Wide-field Infrared Survey Explorer} (\wise). Namely, fluxes in the \emph{W1}, \emph{W2}, \emph{W3}, and \emph{W4} bands have been measured using data from all \wise\ imaging through year 7 of NEOWISE-Reactivation via forced photometry at the locations of LS optical sources and adopting their brightness profiles determined from LS imaging. We retrieved these data from the web service\footnote{\url{https://datalab.noirlab.edu}}. We computed the \emph{g}, \emph{r}, \emph{z}, \emph{W1}, \emph{W2}, \emph{W3}, and \emph{W4} magnitudes in the AB system from the corresponding flux measurements reported (in nanomaggies) in the ls\_dr10.tractor table, using the standard relation $m = 22.5 - 2.5\,\log_{10}{\rm flux}$; additionally converted the \emph{W1}, \emph{W2}, \emph{W3}, and \emph{W4} magnitudes to the Vega system to facilitate comparison with standard MIR AGN diagnostics; and propagated the corresponding uncertainties ($1 / \sqrt{\rm inverse \, variance \, of\, flux}$) accordingly. For sources with negative fluxes in some bands, we estimated 2$\sigma$ upper limits. The LS DR10 fluxes are based on the best-fitting profile model (see Table~\ref{tab:objects}) for each source\footnote{\url{https://www.legacysurvey.org/dr10/description}}. The possible morphological types are point source (`PSF'), round exponential galaxy (`REX'), de Vaucouleurs profile (elliptical galaxy, `DEV'), exponential profile (spiral galaxy, `EXP'), and Sersic profile (`SER'). 

UHS DR2 provides photometric data in two near-infrared (NIR) bands, \emph{J} and \emph{K}, with magnitudes calibrated to the Vega photometric system. We use the Petrosian magnitudes from this catalog, because this type of magnitudes is more appropriate for extended sources. 

Object 1 is \sdss. It is extended in the optical image (Sersic brightness profile), and its optical spectrum is typical of type 2 AGN, as will be discussed in Section~\ref{s:opt_spec}. Furthermore, it has a MIR color $W1-W2\approx 1.3$, which significantly exceeds the standard threshold of $W1-W2=0.8$ for robust AGN identification \citep{stern2012}. If this object is the true counterpart, we can further estimate the ratio of its intrinsic MIR ($\lambda L_\lambda$ at 12\,$\mu$m) and X-ray (2--10\,keV) luminosities. 
Interpolating between the measurements in the \emph{W3} and \emph{W4} bands (virtually unaffected by Galactic absorption),  we estimate $\lambda L_\lambda (12\,\mu{\rm m}) \approx 2.7\times 10^{45}$\,erg\,s$^{-1}$.
For comparison, during the \srg/\art\ all-sky survey, \srgas\ exhibited an unabsorbed luminosity of $6_{-3}^{+6} \times10^{45}$\,erg\,s$^{-1}$ in the rest-frame 2--10\,keV band, whereas during the \art\ pointed observation its X-ray luminosity was significantly lower, $1.0_{-0.3}^{+0.8}\times10^{45}$\,erg\,s$^{-1}$. This implies $\lambda L_\lambda$ (12\,$\mu$m)/$L(2-10\,\rm{keV})=0.3\div 0.9$ and $1.5\div 3.9$ for the `high' and `low' X-ray state, respectively (neglecting possible variability in the MIR). Here, we have corrected the observed X-ray fluxes for intrinsic absorption using $N_{\rm H}\sim 2\times 10^{23}$\,cm$^{-2}$, as inferred from our X-ray spectral analysis, and taken into account the corresponding uncertainty. The derived values are close to the typical 12\,$\mu$m to 2--10\,keV luminosity ratio of $\sim 2$ for AGN \citep{Gandhi2009,Sazonov2012,asmus2015}. Therefore, object 1 is certainly an AGN and is very likely the counterpart of the X-ray source \srgas.

Object 2 is very faint. There is an indication from LS imaging that it is extended (see Table~\ref{tab:objects}) and thus could be a galaxy. The object is not detected in UHS in either the \emph{J} or \emph{K} filter. It has no reported optical spectroscopy and is barely detected by \wise. Its MIR color $W1-W2\lesssim 0.1$ strongly points against an AGN origin. If we nevertheless further consider the hypothesis that it is the counterpart of \srgas\ and is an AGN, then we can estimate the ratio of its intrinsic MIR and X-ray luminosities, similarly to our analysis for object 1 above. 
Specifically, based on the  \emph{W1} measurement and the upper limits on the  \emph{W2},  \emph{W3}, and  \emph{W4} fluxes, and assuming a power-law MIR spectrum (as is typical for AGN, e.g. \citealt{Mullaney2011}), we obtain an upper limit on the $\lambda F_\lambda$ (rest-frame 12\,$\mu$m) flux of $3\times 10^{-15}$\,erg\,s$^{-1}$\,cm$^{-2}$ for redshifts $z<1$, whereas the unabsorbed flux in the rest-frame 2--10\,keV band is $(3\div 14)~\times 10^{-12}$ and $(1.0\div 1.6)~\times 10^{-12}$\,erg\,s$^{-1}$\,cm$^{-2}$ for the high and low X-ray state, respectively (taking into account the uncertainty in the intrinsic X-ray absorption as a function of redshift). Thus, the ratio $\lambda L_\lambda$ (12\,$\mu$m)/$L(2-10\,\rm{keV})$ is less than $3\times 10^{-3}$, much smaller than expected for AGN. We can thus rule out that object 2 is related to \srgas. 

Object 3 is a star at a distance of $660\pm 300$\,pc, with a significant proper motion of $20.5\pm1.0$\,mas\,yr$^{-1}$, according to \gaia\ astrometry (\gaia\ Data Release 3, \citealt{gaiadr3collab}). However, there are strong indications that this is not a single object. First, it is actually listed in LS DR10 as a pair of stellar-like (PSF morphology) objects at an angular distance of 0.95\arcsec\ from each other. 
Secondly, UHS DR2 classifies this source as a single object of morphological type `galaxy'.
Thirdly, the \gaia\ astrometric solution is characterized by a Renormalized Unit Weight Error (RUWE) of 1.85, which significantly exceeds the threshold $\sim 1.3$ for single stars \citep{Castro2024}. Therefore, we might be dealing with a wide binary system. If it were related to \srgas, its unabsorbed X-ray luminosity would be $\sim 10^{32}$\,erg\,s$^{-1}$ (2--10\,keV). Such luminosity could correspond to a cataclysmic variable (CV), which would require the presence of a third (indiscernible) component, namely a white dwarf. However, fitting the joint \art--\xrt\ spectrum by a power-law model absorbed at $z=0$, we find $N_{\rm H} = 7_{-4}^{+9}~\times10^{22}$ (at the 90\% confidence level), whereas CVs never exhibit such strongly absorbed X-ray spectra. Moreover, object 3 is several magnitudes fainter in the optical than typical CVs with such high X-ray fluxes (e.g. \citealt{Sazonov2006}).  

We conclude that \sdss\ (object 1) is the true counterpart of the X-ray source \srgas. It is also worth noting that no optical outbursts have been detected from \sdss\ or any other object within 30\arcsec\ of \srgas\ over the period 2019--2025 according to the Automatic Learning for the Rapid Classification of Events (ALeRCE) light curve classifier \citep{Sanchez2021} of the Zwicky Transient Facility (ZTF) alert stream\footnote{\url{https://alerce.online}}. 

%%%%% 
\subsection{Optical spectrum}
\label{s:opt_spec}

On November 13, 2012 (MJD 56244), a spectrum of SDSS\,J230630.38+155620.4 was obtained by the 2.5-m telescope at the Apache Point Observatory (New Mexico, US) during the Baryon Oscillation Spectroscopic Survey (BOSS, \citealt{boss_survey2013}), part of SDSS. The BOSS spectrograph \citep{boss_spectrograph2013} covers a wavelength range of 3560--10400\,\AA\ with a variable spectral resolution of 2--5\,\AA. The mean full width at half maximum ${\rm FWHM_{res}} = 3.2$\,\AA, combining both the red and blue channels. During the observation, Plate 6134 was mounted in the telescope's focal plane, with Fiber 337 targeting the object. 

\begin{figure*}
    \centering
    \includegraphics{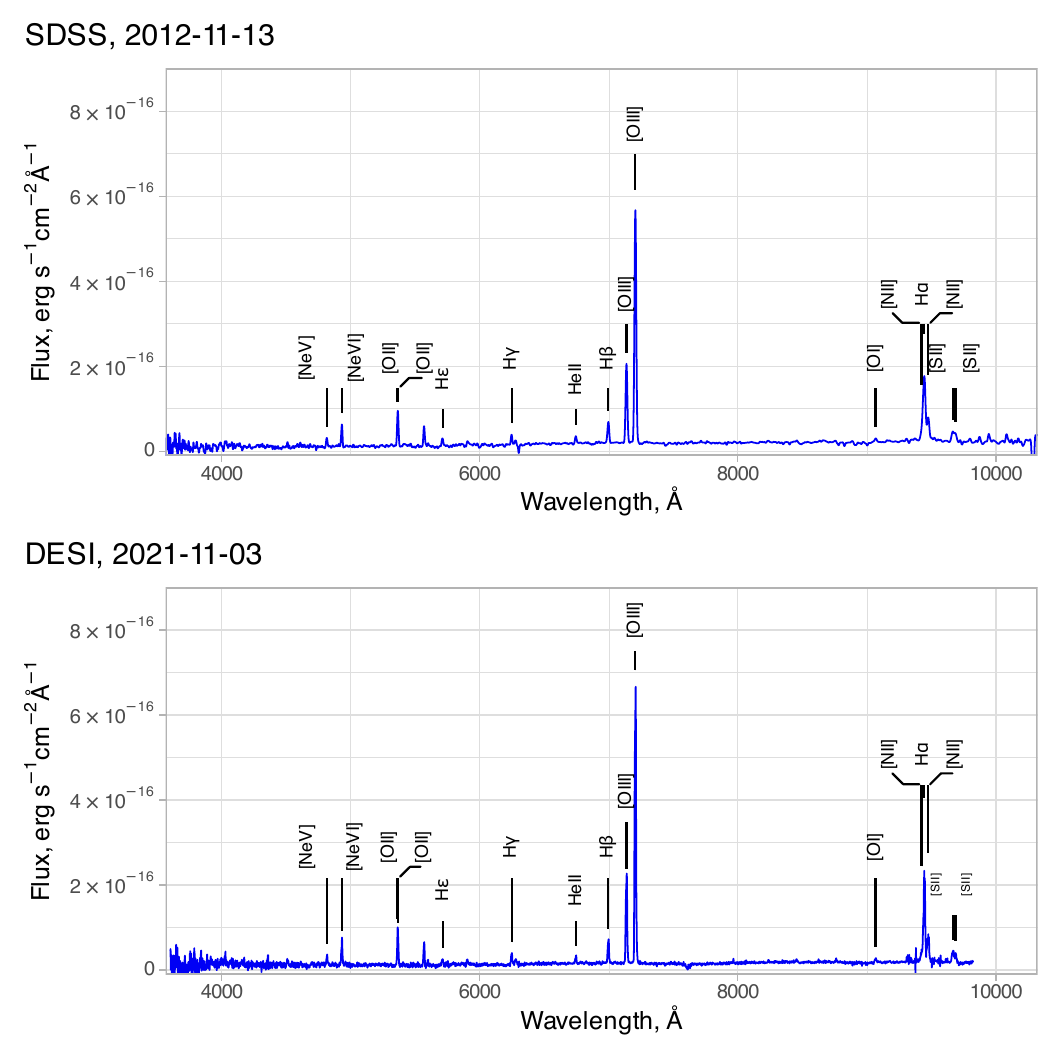}
    \caption{Top panel: SDSS spectrum. Bottom panel: DESI spectrum. Both spectra corrected for Galactic absorption, with emission lines annotated. For visualization purposes, the spectra were smoothed using a moving average algorithm with a window size of five bins.}
    \label{fig:sdss-spec}
\end{figure*}

Figure~\ref{fig:sdss-spec} shows the spectrum of SDSS\,J230630.38+155620.4, retrieved from the SDSS Science Archive Server\footnote{\url{https://dr16.sdss.org/optical}}. We did not apply any additional processing steps to this spectrum. However, the automatic SDSS pipeline (v5\_13\_2) had incorrectly classified the source as `QSO BROADLINE' due to poor spectral line fitting. Therefore, we reconducted the spectral line analysis to obtain correct line measurements. 

We first corrected the spectrum for the Galactic absorption using $E(B-V)=0.20$, the value returned for the source's coordinates by the IRSA Galactic Dust Reddening and Extinction service\footnote{\url{https://irsa.ipac.caltech.edu/applications/DUST}} based on the \citet{schlafly2011} maps. The correction assumes $R_{\rm V}=3.1$ and employs the extinction curve from \citet{cardelli1989}. 
We then fitted Gaussian functions to all identified lines to calculate the redshift. The spectral continuum was fitted by a polynomial function. For convenience, we split the spectrum into several parts: 6200\,\AA\ < $\lambda$ < 7000\,\AA\ (near H$\alpha$), 4200\,\AA\ < $\lambda$ < 5100\,\AA\ (near H$\beta$), and 3317\,\AA\ < $\lambda$ < 4000\,\AA\, excluding the range $\lambda=$3860--3880\,\AA. In each segment, we simultaneously fitted the lines and continuum. After measuring the line centers, we calculated the barycentric-corrected redshift $z=0.43891 \pm 0.00008$ (1$\sigma$ uncertainty). The confidence interval for the redshift was determined using the Student's \emph{t}-distribution. This redshift corresponds to a luminosity distance of $2426.32\pm0.5$\,Mpc for the adopted cosmology.

\subsubsection{Emission lines}
\label{s:opt_lines}

\begin{table}
\centering
\caption{
\label{tab:lines}
Spectral line characteristics of SDSS\,J230630.38+155620.4. 
}
\centering
\setlength{\tabcolsep}{1pt}
\begin{tabular}[t]{lcccc}
\toprule
Line & Wavelength, \AA & \CellWithForceBreak{Flux, $10^{-15}$  \\ erg\,s$^{-1}$\,cm$^{-2}$}  & EW,\,\AA & \CellWithForceBreak{FWHM, \\ $10^2$\,km\,s$^{-1}$}\\
\midrule
{}[NeV]$\lambda$3346 & $4814.0\pm0.3$ & $0.269\pm0.018$ & $-17.8\pm1.1$ & $6.1\pm0.7$\\
{}[NeVI]$\lambda$3426 & $4930.83\pm0.12$ & $0.601\pm0.017$ & $-37.2\pm1.0$ & $5.1\pm0.2$\\
{}[OII]$\lambda$3726 & $5364.33\pm0.08$ & $1.100\pm0.019$ & $-61.4\pm1.0$ & $5.45\pm0.15$\\
H$\epsilon$ & $5710.4\pm0.5$ & $0.29\pm0.02$ & $-17.3\pm1.3$ & $6.3\pm0.8$\\
H$\gamma$ & $6246.0\pm0.3$ & $0.302\pm0.019$ & $-13.2\pm0.8$ & $4.0\pm0.4$\\
HeII & $6743.9\pm0.3$ & $0.251\pm0.014$ & $-9.5\pm0.5$ & $4.4\pm0.4$\\
H$\beta$ & $6996.02\pm0.12$ & $0.812\pm0.017$ & $-28.6\pm0.5$ & $5.01\pm0.17$\\
{}[OIII]$\lambda$4959 & $7136.57\pm0.04$ & $3.10\pm0.02$ & $-105.5\pm0.8$ & $4.90\pm0.06$\\
{}[OIII]$\lambda$5007 & $7205.33\pm0.02$ & $9.40\pm0.05$ & $-314.4\pm1.6$ & $4.78\pm0.03$\\
{}[OI]$\lambda$6300 & $9068.6\pm0.8$ & $0.154\pm0.019$ & $-4.9\pm0.5$ & $4.2\pm0.9$\\
{}[NII]$\lambda$6548 & $9421.3\pm1.0$ & $0.41\pm0.05$ & $-12.8\pm1.6$ & $4.4\pm1.0$\\
H$\alpha$ & $9443.90\pm0.19$ & $4.11\pm0.10$ & $-124\pm3$ & $5.4\pm0.2$\\
{}[NII]$\lambda$6584 & $9474.5\pm0.4$ & $1.22\pm0.06$ & $-36.9\pm1.6$ & $4.8\pm0.4$\\
{}[SII]$\lambda$6716 & $9664.5\pm1.2$ & $0.47\pm0.07$ & $-13.8\pm1.9$ & $4.8\pm1.0$\\
{}[SII]$\lambda$6731 & $9686.5\pm1.6$ & $0.40\pm0.09$ & $-12\pm3$ & $5.1\pm1.8$\\
\bottomrule
\end{tabular}
\begin{flushleft}
The wavelengths are presented in the observer's frame, while the equivalent widths in the source rest frame. The uncertainties are given at the 1$\sigma$ confidence level. 
\end{flushleft}
\end{table}

The identified lines, along with their observed wavelengths, fluxes, equivalent widths (EW), and physical widths (FWHM) are presented in Table~\ref{tab:lines}.
The ${\rm FWHM}$ values (converted to velocities) were calculated as: 
\begin{equation}
    {\rm FWHM}=\frac{\sqrt{{\rm FWHM}^2_{\rm mes} - {\rm FWHM}^2_{\rm res}(\lambda)}}{1+z},
\end{equation}
where ${\rm FWHM}_{\rm mes}$ is the measured line width and ${\rm FWHM_{res} (\lambda)}$ is the spectral resolution at the observed central wavelength $\lambda$. The confidence intervals for EW were obtained using Monte Carlo simulations. Specifically, assuming that the flux errors obey a normal distribution, we generated 1,000 spectral realizations and estimated EW for each of them. The confidence intervals were then estimated from the derived EW distribution.
%Table~\ref{tab:lines} presents the measured line characteristics. 

\begin{figure}
\centering
\includegraphics[width=1\columnwidth]{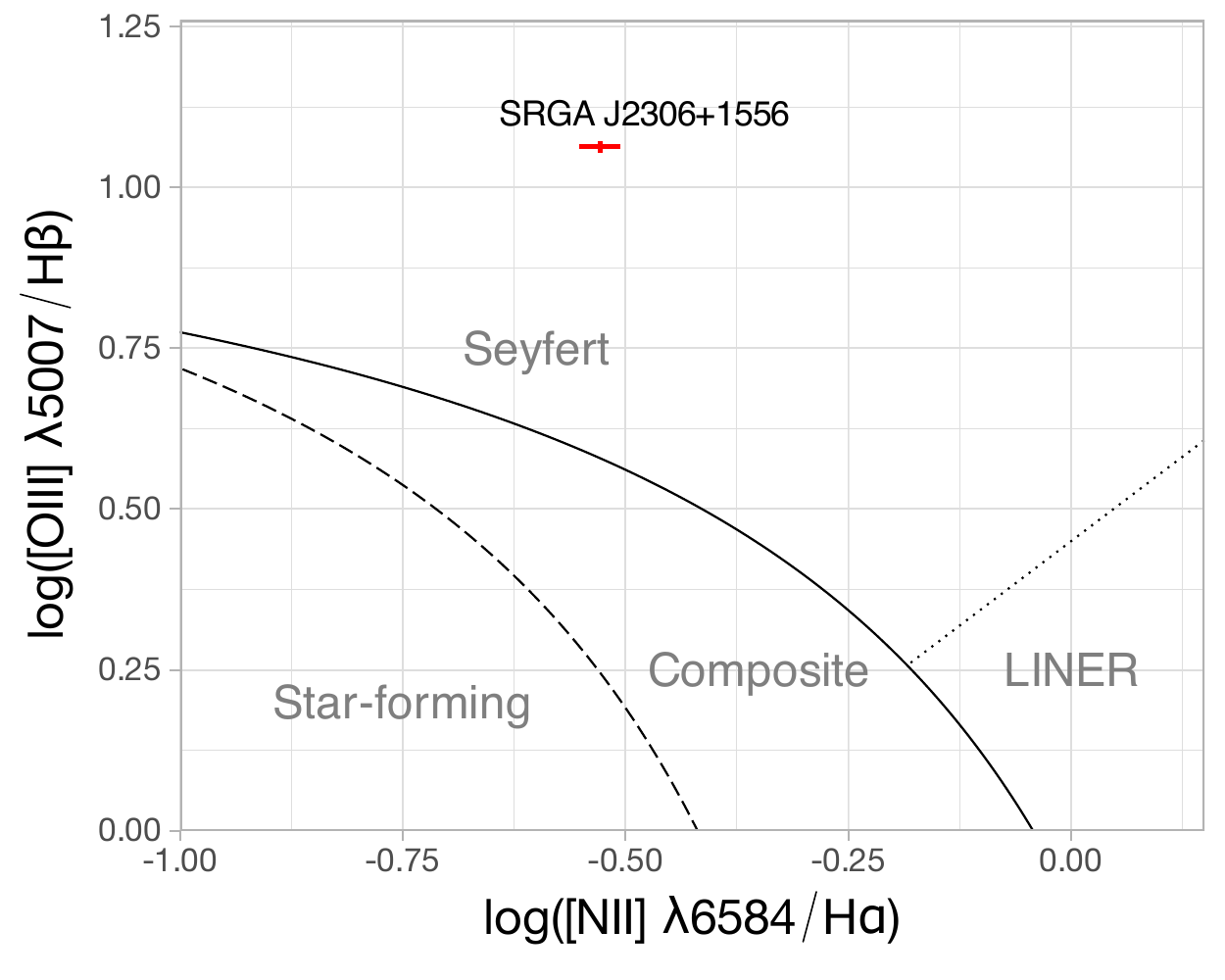}
\caption{
Location of \srgas\ on the BPT diagram \citep{bpt} and the corresponding 1$\sigma$ error bars (in red). The boundaries separating different types of galaxies are adopted from: \protect\cite{kauff03}, solid line; \protect\cite{kewly01}, dashed line; and \protect\cite{scha07}, dotted line.}
\label{fig:bpt}
\end{figure}

All the emission lines detected in the spectrum, namely those of the Balmer series and many forbidden lines, have Doppler widths of approximately 500\,km\,s$^{-1}$, which is typical of type 2 AGN. The flux ratios of the key pairs of lines are \othreehb\,=\,$1.064\pm0.009$ and \ntwoha\,=\,$-0.53\pm0.02$, thus the source falls within the Seyfert region of the Baldwin, Phillips \& Terlevich (BPT) diagram (Fig.~\ref{fig:bpt}). 

Based on the measured fluxes of the Balmer lines, we can estimate the Balmer decrement in the direction of the narrow-line region (NLR): $({\rm H}\alpha/{\rm H}\beta)_{\rm measured} = 5.1\pm 0.2$. This value falls into the 1$\sigma$ region (3.5--5.5) of the distribution of NLR Balmer decrements found by \cite{lu2019} for a large sample of AGN from the SDSS survey and implies the presence of dust in the NLR and/or the surrounding medium. 
If there were no dust, the intrinsic NLR Balmer decrement is expected to be H$\alpha$/H$\beta\approx 3.1$, i.e. slightly higher than the value 2.85 corresponding to case B recombination, which is attributed to collisional excitation of H$\alpha$ \citep{osterbrock&ferland2006}.  
We can then estimate the intrinsic extinction in the direction of \srgas's NLR (again assuming $R_V=3.1$ and the extinction curve of \citealt{cardelli1989}): $A_V = 7.202 [\log({\rm H}\alpha/{\rm H}\beta)_{\rm measured} - \log 3.1]=1.5\pm0.1$\,mag, or $E(B-V)=0.50\pm0.03$ (here, the $B$ and $V$  bands are defined in the object's rest frame). Accordingly, the extinction in the O[III]$\lambda$5007 emission line, which will be discussed below, can be estimated as $A(O[III]) = 3.475 A_V / R_V=1.7 \pm 0.1$\,mag.

\subsubsection{Continuum}
\label{s:opt_cont}

The spectrum exhibits a weak but significant optical continuum. The [NeV]$\lambda$3346, [NeVI]$\lambda$3426, and [OII]$\lambda$3726 emission lines contribute $(1.97 \pm 0.03)\times 10^{-15}$\,erg\,s$^{-1}$\,cm$^{-2}$, or 10\% of the total emission in the \emph{g} band of the Dark Energy Camera (DECam). The H$\epsilon$, H$\gamma$, HeII, H$\beta$, [OIII]$\lambda$4959, and [OIII]$\lambda$5007 lines contribute $(1.415 \pm 0.007)\times 10^{-14}$\,erg\,s$^{-1}$\,cm$^{-2}$, or 40\% to the \emph{r} band. The [OI]$\lambda$6300, [NII]$\lambda$6548, H$\alpha$, [NII]$\lambda$6584, [SII]$\lambda$6716, and [SII]$\lambda$6731 lines contribute $(6.8 \pm 0.2)\times 10^{-15}$\,erg\,s$^{-1}$\,cm$^{-2}$, or 14\% to the \emph{z} band. 

Correcting the visual/IR photometric magnitudes quoted in Table~\ref{tab:lines} for the Galactic extinction 
in the direction of \sdss\
following the standard procedure described in the LS DR10 documentation, we obtain total extinction-corrected magnitudes of $g=20.43 \pm 0.02$, $r=19.358 \pm 0.009$, $z = 18.510 \pm 0.015$, $J=17.6 \pm 0.2$ and $K=15.77 \pm 0.12$, $W1=14.563 \pm 0.008$, $W2=13.284 \pm 0.009$, $W3=9.27 \pm 0.02$, and $W4=5.78 \pm 0.06$. If we now assume that line emission does not contribute significantly outside the \sdss\ central region of $\sim 1\arcsec$, or $\sim 6$\,kpc radius that corresponds to the SDSS/BOSS spectroscopy aperture, then subtracting the line fluxes estimated from spectrophotometry above, we derive Galactic extinction-corrected magnitudes of $g = 20.54 \pm 0.02$, $r = 19.918 \pm 0.018$, and $z = 18.67 \pm 0.02$ for the optical continuum emission. We will use these values, as well as the $J$ and $K$ magnitudes quoted above, to estimate the mass of the \sdss\ host galaxy (in Sect.~\ref{s:host}). 

The continuum luminosity at rest-frame 5100\AA\ (a quantity often used in studies of galaxies and AGN), averaged over five spectral bins around this wavelength, is $\lambda L_\lambda = (9.9\pm0.7)\times 10^{43}$\,erg\,s$^{-1}$. 
In summary, based on the totality of optical line and continuum properties, in combination with the high bolometric luminosity of \srgas=\sdss\ (see below), we can confidently classify it as a type 2 quasar. 

When we were completing this paper, the first release of data from the Dark Energy Spectroscopic Instrument (DESI DR1) came out \citep{desi2025}. It provided a new spectrum of \sdss, which was taken on November 3, 2021 (with an exposure time of 1360\,seconds), i.e about nine years after the SDSS observation. This new spectrum is shown in the lower panel of Fig.~\ref{fig:sdss-spec}. DESI \citep{desi_instrum2022}) is installed on a 4-m telescope and employs three spectral cameras that operate in the $B$, $R$, and $Z$ bands and cover the wavelength range from 3600 to 9800\,\AA; the spectral resolution ($\lambda/\Delta\lambda$) varies from approximately 2000 to 5000 depending on the wavelength, that is typically a factor of two better than SDSS. The fluxes of emission lines in the SDSS spectrum of \sdss\ are systematically $\sim 15$\% higher than in the DESI spectrum, and the continua show the same $\sim 15$\% offset. 
This suggests that the difference arises from absolute flux calibration uncertainties rather than intrinsic variability of the source. The absence of substantial spectral variability in the optical band on a rest-frame time scale of $\sim 6$\,years is unsurprising, given that \sdss\ is a type 2 quasar.

%%%%% 
\subsection{Radio data}
\label{s:radio_data}

\begin{figure}
    \centering
    \includegraphics[angle=-90,width=1.2\columnwidth]{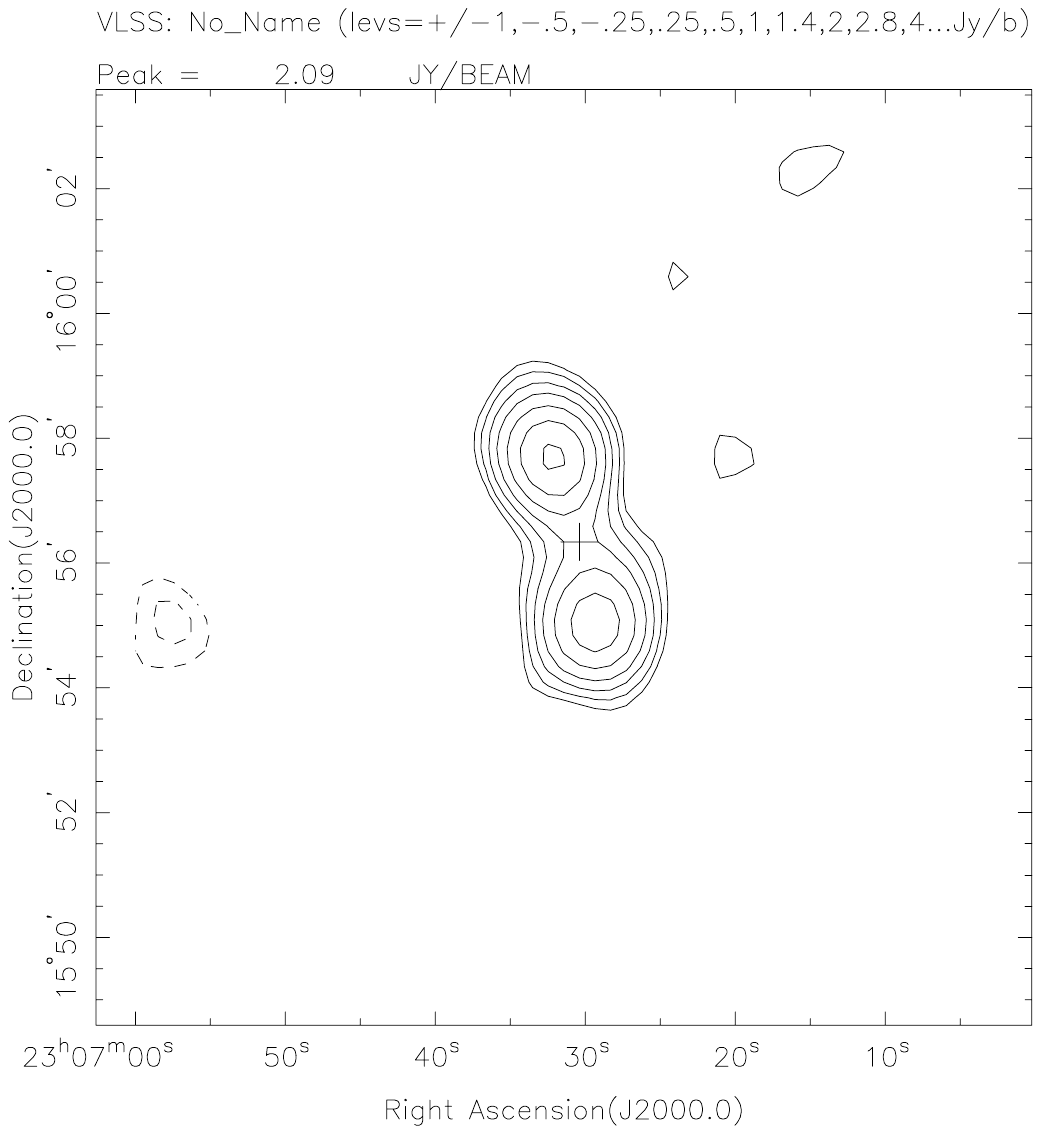}
    \includegraphics[angle=-90,width=1.2\columnwidth]{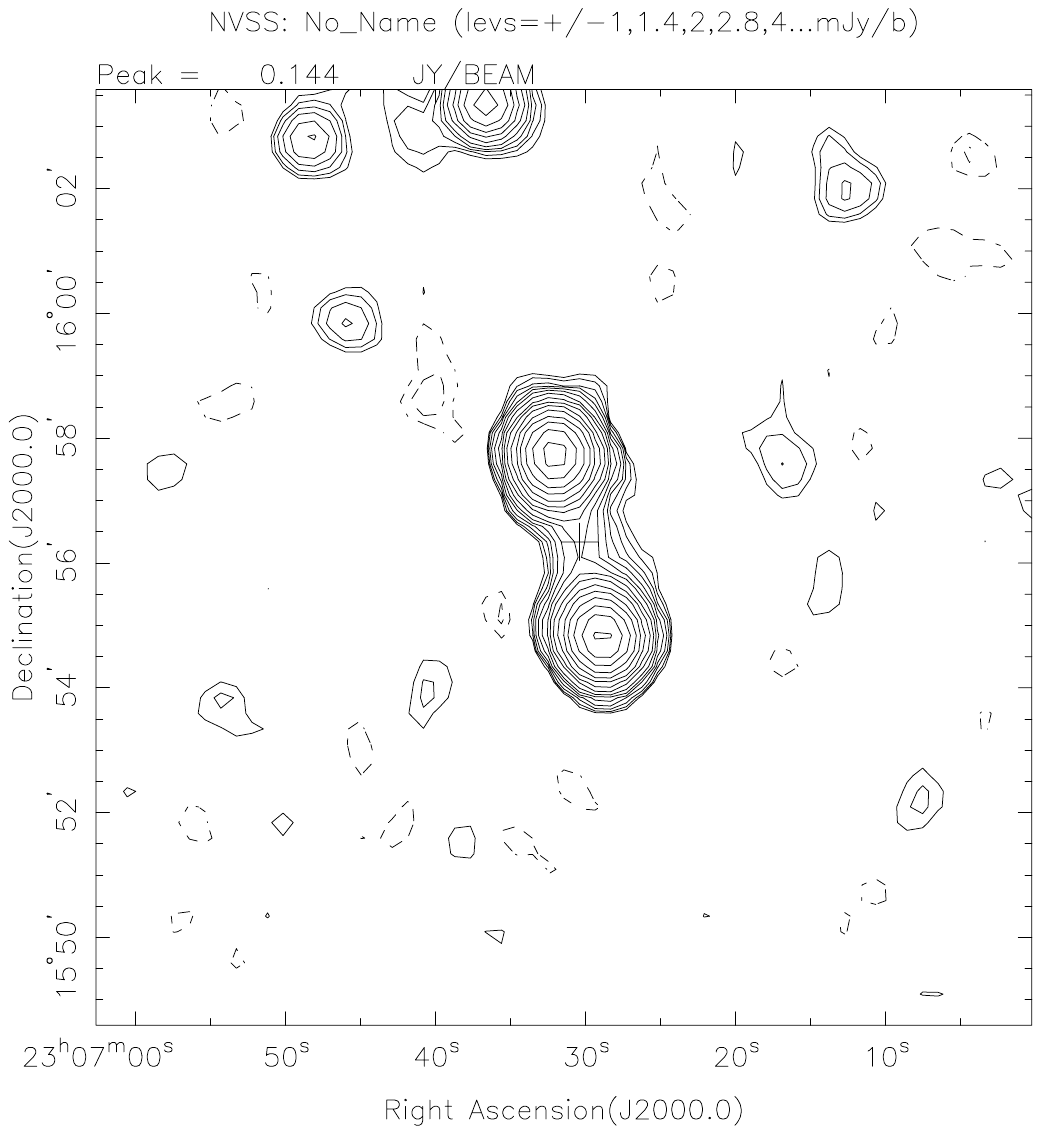}
    \caption{
        Radio maps of \srgas\ at 74\,MHz (VLSSr, top panel) and 1400\,MHz (NVSS, bottom panel). The cross indicates the position of the optical counterpart of \srgas.
    }
    \label{fig:rad_maps}
\end{figure}

It turns out that \srgas\ = \sdss\ has been known as a radio source for a long time (see Table~\ref{tab:radio_surv} for key characteristics of the surveys discussed below). The source, roughly located at 156\arcsec\ from the quasar's optical position, was first detected in the 4C survey at 178\,MHz \citep{1967MmRAS..71...49G}. Its measured flux density was 2.2\,Jy. Later observations with better angular resolution revealed a double radio source near \srgas\ at 74\,MHz (VLSSr survey, \citealt{2014MNRAS.440..327L}, see Fig.~\ref{fig:rad_maps}, top panel\footnote{The radio map is taken from \url{https://www.cv.nrao.edu/vlss/VLSSpostage.shtml}}) and 150\,MHz (TGSS survey, \citealt{Intema2017}). The bright components have almost equal flux density: $3.02\pm0.41$ and $2.79\pm0.39$\,Jy at 74\,MHz; $1.51\pm0.15$ and $1.32\pm0.13$\,Jy at 150\,MHz for the northern and southern components, respectively. The corresponding spectral indices of these components $\alpha_{74-150}=-0.98\pm0.24$ and $-1.06\pm0.24$, which implies optically thin, aged synchrotron emission. The RACS survey (887.5\,MHz) also reveals a double radio structure, with the total flux density of about 640\,mJy and two brightest Gaussian components of 137 and 120\,mJy \citep{2021PASA...38...58H}. 

\begin{table}
    \centering
    \begin{tabular}{c|c|c|c|c}
    \toprule
      Survey   & Frequency, & Angular & Period of & Reference \\
       & MHz & resolution & observations & \\
      \midrule
       VLSSr  & 74 & 80\arcsec & 2001--07 & \cite{2014MNRAS.440..327L} \\
       TGSS  & 150 & 25\arcsec & 2010--12 & \cite{Intema2017} \\
       4C  & 178 & 3.5\arcmin & 1960-65 & \cite{1967MmRAS..71...49G} \\
       TXS  & 365 & 20\arcsec & 1974--83 & \cite{1996AJ....111.1945D} \\
       MRC  & 408 & 3\arcmin & 1968--78 & \cite{1991Obs...111...72L} \\
       RACS  & 887.5 & 15\arcsec & 2019 & \cite{2021PASA...38...58H} \\
       NVSS  & 1400 & 45\arcsec & 1993--97 & \cite{1998AJ....115.1693C} \\
       VLASS  & 3000 & 2.5\arcsec & 2017--23 & \cite{2020PASP..132c5001L} \\
       87GB  & 4850 & 3.5\arcmin & 1987 & \cite{1991ApJS...75.1011G} \\
       GB6 & 4850 & 3.5\arcmin & 1986--87 & \cite{1996ApJS..103..427G} \\
       \bottomrule
    \end{tabular}
    \caption{Radio surveys discussed in the text.}
    \label{tab:radio_surv}
\end{table}

\begin{figure}
    \centering
    \includegraphics[width=1\columnwidth]{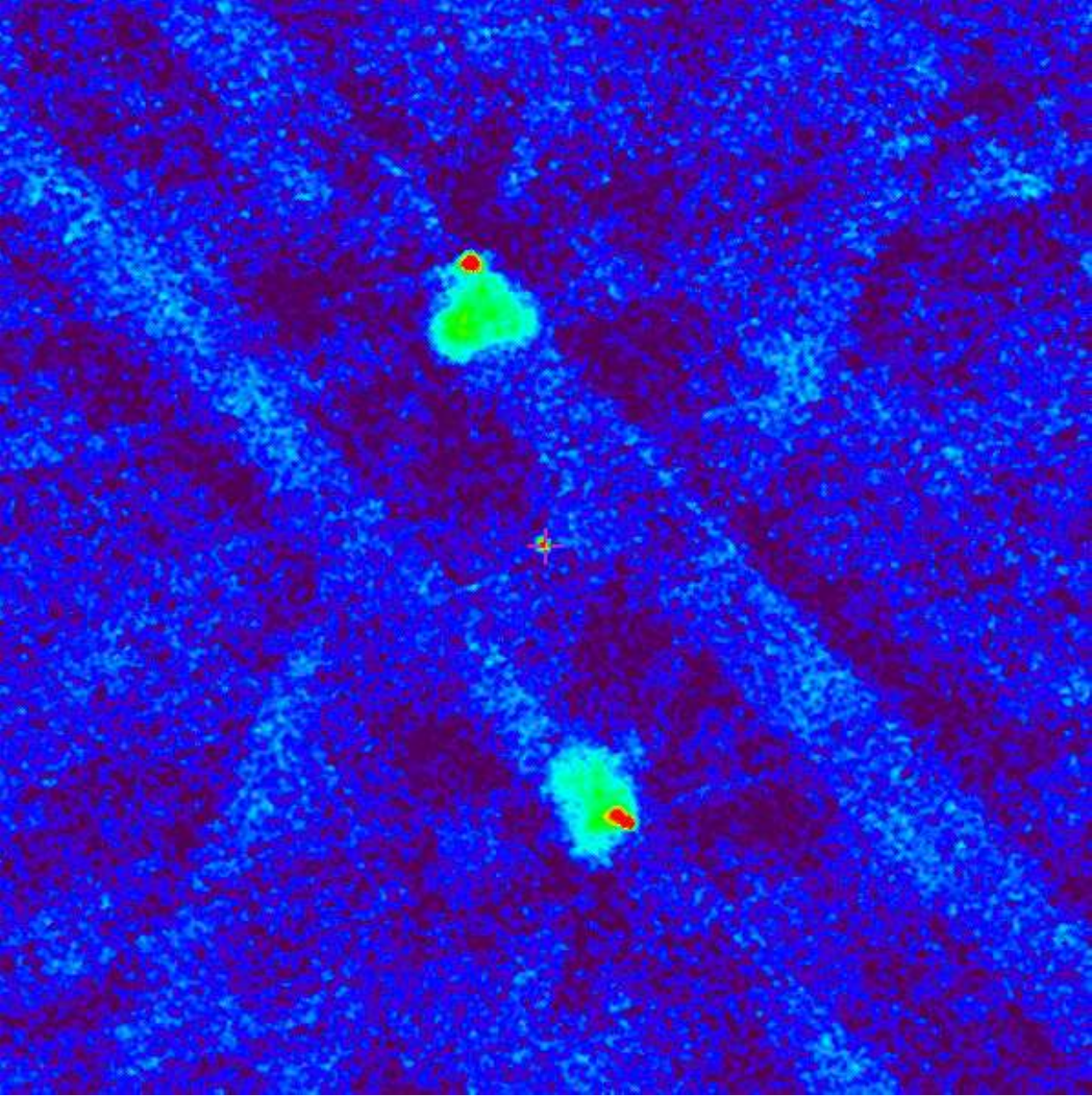}
    \caption{
    Radio image (6.5\arcmin\ on a side) of \srgas\ at 3000\,MHz based on median stacks of the Quick Look images from the three VLASS epochs. The red cross indicates the position of the optical counterpart of \srgas. The northern (red point at the top of the image) and southern (two red points at the bottom of the image) components are hotspots in extended radio lobes (denoted with green colors).}
    \label{fig:img-radio_stack}
\end{figure}

The NVSS morphology at 1400\,MHz shows two bright components (Fig.~\ref{fig:rad_maps}, bottom panel\footnote{The radio map is taken from \url{https://www.cv.nrao.edu/nvss/postage.shtml}}) with approximately equal flux density for the northern and southern components: $215\pm8$ and $193\pm7$\,mJy, respectively \citep{1998AJ....115.1693C}. However, there is also a weak central component ($14\pm1$\,mJy) at 5\arcsec\ from \sdss. The triple morphology with a central component ($\approx5$\,mJy) is confirmed at 3000\,MHz by the Very Large Array Sky Survey (VLASS) (Fig.~\ref{fig:img-radio_stack}, \citealt{2020PASP..132c5001L}). This central component coincides with the optical position of the quasar within 0.3\arcsec. The central component does not show variability according to VLASS data: $4.81\pm0.28$\,mJy (Epoch 1, October 2017) and $4.64\pm0.39$\,mJy (Epoch 2, August 2020). 
The northern component (N) has a flux density of $20.47\pm0.69$\,mJy in Epoch 1 and $25.90\pm1.12$\,mJy in Epoch 2. The southern component (S) appears as double: $13.00\pm0.26$\,mJy ($\text S_{\text a}$) and $11.88\pm0.66$\,mJy ($\text S_{\text b}$) in Epoch 1 (the separation between these subcomponents is 4.7\arcsec); $12.81\pm0.36$\,mJy ($\text S_{\text a}$) and $9.69\pm0.82$ mJy ($\text S_{\text b}$) in Epoch 2. The region near \srgas\ was also observed during Epoch 3 of the VLASS survey in January 2023, but a catalog of radio sources for Epoch 3 is not yet available.  

At yet higher frequencies, there are only individual measurements at 4.85\,GHz with the Green-Bank 91m radio telescope in 1986--87 \citep{1991ApJS...75.1011G,1991ApJS...75....1B,1996ApJS..103..427G}. However, these observations were made with 3.5\arcmin\ angular resolution and the radio source was not resolved, with the total flux density of about 110\,mJy. 

\begin{figure}
    \centering
    \includegraphics[width=1\columnwidth]{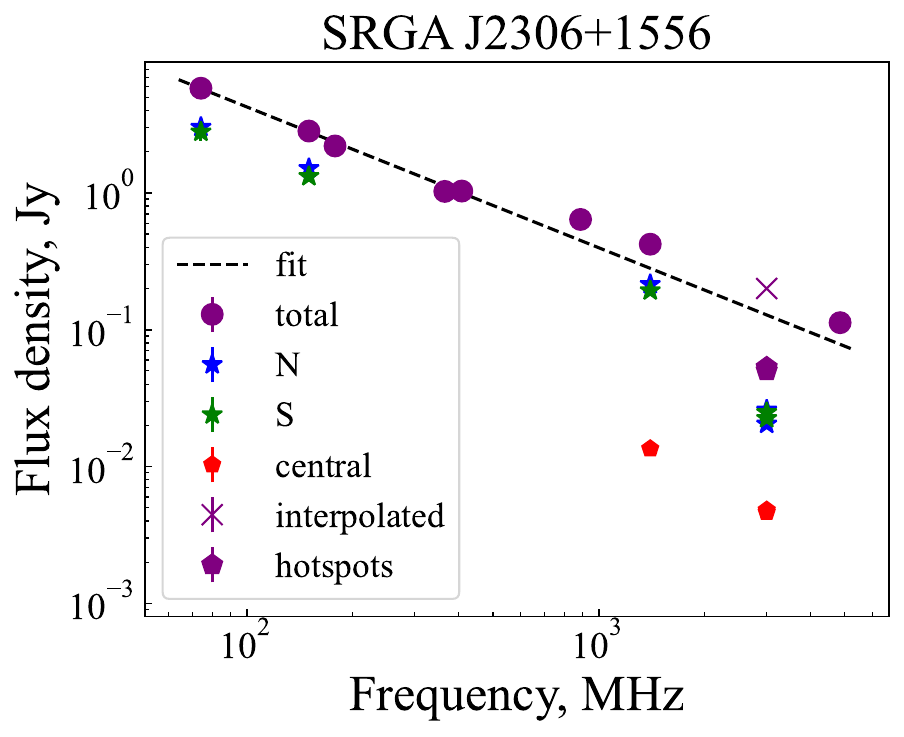}
    \caption{
        Broadband spectrum of the radio counterpart of \srgas. The purple circles denote the total flux density of all components; the blue and green stars denote the flux density of the northern and southern components, respectively; the red pentagons show the flux density of the central component. The purple cross indicates the interpolated flux density at 3000\,MHz from the measurements at 888--4850\,MHz. The purple pentagons denote the flux density from the hotspots at 3000\,Mhz.
        The dashed line represents a power-law model $F_\nu \sim \nu^{-1}$ for the total flux density.
    }
    \label{fig:rad_spec}
\end{figure}

Figure~\ref{fig:rad_spec} presents the spectra of the northern, southern and central components as well as the total spectrum of the radio counterpart of \srgas. The total spectrum can be well described by a steep power law, $F_\nu\sim\nu^{\alpha}$ with $\alpha\approx-1$, despite the fact that measurements were made over a period of 55 years. This indicates that the source is not strongly variable in the radio band. 

One exception to this power law is the total measured flux density of 50--55\,mJy at 3\,GHz: this is much less than the expected flux density of about 200\,mJy based on the steep power-law spectrum inferred from measurements at lower and higher frequencies. This problem can be solved by accounting for the presence of diffuse radio emission at 3\,GHz. VLASS has the highest angular resolution among the radio surveys discussed here and detects only the brightest components with high signal-to-noise ratio. A significant fraction of radio emission at 3\,GHz is spread over an area that is bigger than the VLASS beam size. Let us roughly estimate the potential contribution of the diffuse emission to the total flux density. The RMS sensitivity per epoch is about 0.12\,mJy/beam for VLASS and the beam size is about 3\arcsec. The separation between the VLASS N and S components is about 210\arcsec, and the RMS is 147\,mJy per circle of 105\arcsec\ radius. We thus obtain an upper limit of about 200\,mJy for the total flux density at 3\,GHz by summing the flux density of the three VLASS components and this RMS value. We can further assume that the extended morphology at 3\,GHz is dumbbell-shaped like that at 1.4\,GHz. Then we can consider the N and S components as circles of radius $r=xR$, where $R=105\arcsec$ is the angular separation between each of these components and the central one. It is likely that $x=$0.6--0.8, thus $r=63$--84\arcsec. The RMS is about 53\,mJy per circle of 63\arcsec\ radius, whereas the RMS for the bridge between the N and S circles is about 24\,mJy for $r=63\arcsec$. In total we get an RMS of about 130\,mJy for the area of the radio source. Hence, the upper limit for the total flux density in the considered case is about 180\,mJy, in satisfactory agreement with the power-law spectrum shown in Fig.~\ref{fig:rad_spec}.

We suppose that the VLASS N and S components are hotspots. This is confirmed by the interactive radio image based on median stacks of the Quick Look images from all three VLASS epochs\footnote{Taken from \url{https://archive-new.nrao.edu/vlass/HiPS/All\_VLASS/Quicklook}} of the region centered on \sdss\ (Fig.~\ref{fig:img-radio_stack}). This stacked image demonstrates the presence of radio lobes that are not detected in individual epochs.

The radio core contribution can be characterized as the ratio of the flux density of the central component and the total one: $C_{\nu}=S_{\nu,\rm c}/S_{\nu,\rm total}$. We estimate 
$C_{1.4\rm \,GHz}=0.032\pm0.04$ and $C_{3.0\rm \,GHz}=0.025\pm0.001$ (if we take $S_{3.0,\rm total}=180\div200$\,mJy), thus the contribution of the radio core does not become more important at higher frequencies. 

The projected distance between the centers of the symmetric components is about 0.9\,Mpc according to the VLSSr and TGSS data, and 1.0\,Mpc according to NVSS (at redshift $z=0.4389$). The projected distance between the VLASS components is 1.2\,Mpc, confirming that these are hotspots. The central VLASS component has the deconvolved size $\leq0.9\times0.5$\arcsec (the projected size is $\leq5\times3$\,kpc), which suggests its association with the central region of the quasar's host galaxy. These estimates imply that the radio counterpart of \srgas\ is a giant radio galaxy. Its radio power is $P_{1400\,\rm MHz} = 2.86\times10^{26}$\,W\,Hz$^{-1}$, which is equivalent to $\nu L_{\nu}=4\times10^{42}$\,erg\,s$^{-1}$. This high radio power and the morphological features at 3\,GHz discussed above mean that this giant radio galaxy belongs to FR\,II type. Our conclusion about the nature of the radio counterpart of \srga\ confirms the findings of \cite{2024ApJS..273...30B} who reported the discovery of J2306+1556 as a giant radio quasar among 34 new giant radio sources based on TGSS data.

%%%%%%%%%%%%%%%%%%%%%%%%%%%%%%%%%%%%%%
\section{Spectral energy distribution}
\label{s:sed}
%%%%%%%%%%%%%%%%%%%%%%%%%%%%%%%%%%%%%%

\begin{figure}
    \centering
    \includegraphics[width=\columnwidth]{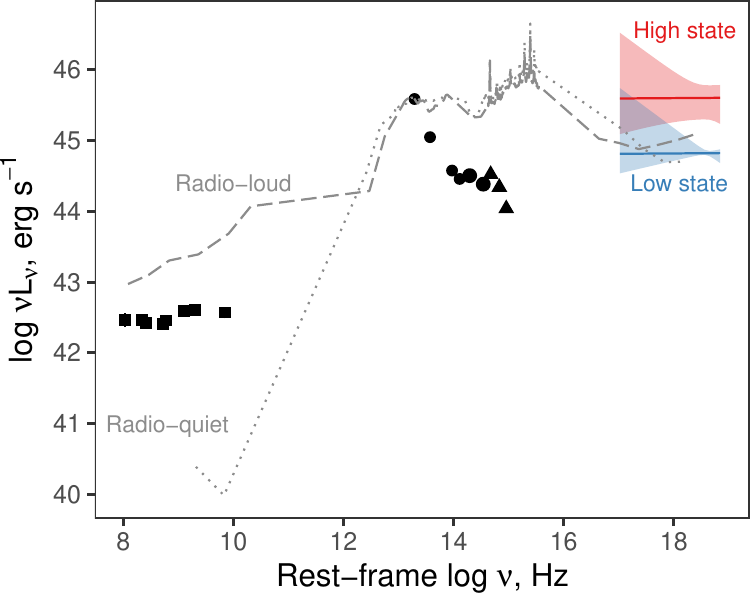}
    \caption{
Spectral energy distribution of \srgas. The squares, circles, and triangles show the radio (74.0--4850\,MHz), IR (\emph{W1}, \emph{W2}, \emph{W3}, \emph{W4}, \emph{J}, \emph{K}), and optical (\emph{grz}) measurements, respectively, corrected for Galactic extinction and converted to intrinsic luminosity densities.  
In the X-ray band (rest-frame 0.4--29\,keV), the blue and red solid lines show the intrinsic power-law component for the `low' and `high' source states, while the surrounding shaded regions provide the 1$\sigma$ confidence intervals. The dotted and dashed lines show the radio-quiet and radio-loud type 1 quasar templates from \citet{shang2011}, each normalized to the $W4$ measurement. The optical and infrared photometric measurements, except for \emph{W4}, lie significantly below the spectral templates because the continuum emission of \srgas\ is heavily obscured by the dusty torus, as expected for a type 2 quasar, while the templates represent unobscured type 1 quasars.
}
    \label{fig:sed}
\end{figure}

We constructed the broadband spectral energy distribution (SED) of \srgas\ from radio to X-ray energies (Fig.~\ref{fig:sed}). Presumably, the radio band represents synchrotron radiation from the jets and lobes, the infrared band reveals emission from the accretion disc reprocessed by the dusty torus, the optical emission mostly comes from the host galaxy, and the X-rays originate in the hot corona of the accretion disc. 

In the radio band, we used the total flux measurements in the observed range of 74.0--4850\,MHz, excluding the 3\,GHz high-angular-resolution measurement by VLASS because it underestimates the diffuse radio emission, as was discussed in Section~\ref{s:radio_data}. For the optical and infrared bands, we used the photometry from DESI LS DR10 (\emph{grz}, \emph{W1}, \emph{W2}, \emph{W3}, \emph{W4}) and UKIRT HS DR2 (\emph{JK}), corrected for Galactic extinction (see Sect.~\ref{s:opt_cont}). The emission-line contribution was not subtracted. 
In the X-ray band (0.3--20\,keV in the observed frame, corresponding to 0.4--29\,keV in the rest frame), for the source's low state we used the power-law spectral model with a free slope (see Table~\ref{tab:xray-params}), corrected for Galactic and intrinsic absorption, and evaluated the corresponding confidence intervals using a Markov chain Monte Carlo technique (the Goodman-Weare algorithm) with 30 walkers and a total of 600,000 steps.
For the high state (for which we do not have X-ray spectral information), we assumed the same spectral shape as for the low state and additionally took the flux measurement uncertainty into account. 

It is interesting to compare the SED of \srgas\ with the composite SEDs of radio-loud and radio-quiet type 1 quasars from \citet{shang2011}, which span a range from rest-frame 2.45\,m (radio) to 11.6\,keV (X-ray). These templates have been obtained by averaging the SEDs of many objects and are thus representative of the corresponding populations (although there is a significant scatter from object to object). The object of our study is a type 2 quasar, whose intrinsic emission from the NIR through the soft X-ray band is expected to be nearly totally obscured from us by the dusty torus. We thus normalize the \citet{shang2011} templates to the luminosity density of \srgas\ in the $W4$ band, i.e. at a rest-frame wavelength of 15\,$\mu$m, where the torus is likely to be optically thin to its own radiation so that the observed luminosity should not depend strongly on the viewing angle. We can clearly see from Fig.~\ref{fig:sed} that \srgas\ is a radio-loud quasar, since the contribution of radio emission to its total radiative output is much closer to the radio-loud template than to the radio-quiet one. 

We can estimate the intrinsic bolometric luminosity of \srgas\ by integrating the radio-loud type 1 SED template. Since the IR `bump' (centered at around 15\,$\mu$m) is presumed to be reprocessed optical--ultraviolet (UV) emission from the accretion disk and soft X-ray emission from its hot corona, it is reasonable to integrate the SED only at wavelengths shorter than 1\,$\mu$m.
The dominant contribution is provided by the `Big Blue Bump' (1\,$\mu$m -- 1\,keV): $L_{\rm BBB} = 5\times10^{46}$\,erg\,s$^{-1}$ (including the significant contribution of emission lines), and the X-ray (1--11.6\,keV) component provides an additional minor contribution of $2.6\times10^{45}$\,erg\,s$^{-1}$. 
For comparison, the integration of the radio-loud template over the IR band, namely over the wavelengths longer than 1\,$\mu$m, yields: $L_{\rm >1\mu m} = 2.1\times10^{46}$\,erg\,s$^{-1}$, i.e. $\sim 40$\% of the BBB luminosity, which probably reflects the covering fraction of the obscuring torus as seen from the accretion disk.

The actually measured unabsorbed luminosity of \srgas\ in the 2--10\,keV band is $\Lx=1.0^{+0.8}_{-0.3}\times 10^{45}$ and $6^{+6}_{-3}\times10^{45}$\,erg\,s$^{-1}$ in its low and high X-ray state, respectively. 

Assuming a power law with $\Gamma=1.8$ and an exponential-cutoff energy $E_{\rm cut}=150$\,keV, as typical of AGN \citep{Malizia2014,Ricci2017}, we can estimate the total luminosity of the X-ray continuum between 1 and 1000\,keV ($\Lhx$) at $\sim 4\times 10^{45}$ and $\sim 2\times 10^{46}$\,erg\,s$^{-1}$ for the low and high state, respectively. Consequently, a more realistic estimate of the quasar's bolometric luminosity is $\Lbol=L_{\rm BBB}+\Lhx\sim 5.4\times 10^{46}$ and $\sim 7\times 10^{46}$\,erg\,s$^{-1}$ for the low and high state, respectively, and the corresponding ratios $\Lhx/\Lbol\sim 0.07$ and $\sim 0.3$. For comparison, the summed emission of all quasars in the Universe is characterized by an average value $\Lhx/\Lbol\sim 0.13$ \citep{Sazonov2004}. Neglecting the uncertainties, we can thus regard the low and high states of \srgas\ as `X-ray dim' and `X-ray bright', respectively. However, there is of course a possibility that the UV luminosity, about which we have no direct information, also increased during the X-ray outburst. 

While the X-ray observations presented here have revealed that \srgas\ is significantly variable on time scales of a few years, its typical level of activity on much longer times scales can be inferred from observations of emission from the (kpc-scale) NLR. Using the measured flux in the {}[OIII]$\lambda$5007 line and the corresponding intrinsic absorption correction determined in Section~\ref{s:opt_lines},
we can estimate the luminosity in this line at $L_{\rm O[III]} = (3.2 \pm 0.3) \times 10^{43}$\,erg\,s$^{-1}$. The corresponding ratio to the 2--10\,keV luminosity $L_{\rm O[III]} / \Lx \approx 0.03 \pm 0.02$ and $L_{\rm O[III]} / \Lx \approx 0.005 \pm 0.004$ for the low and high X-ray state, respectively. This can be compared with a typical value $\sim 0.01$ for AGN (e.g. \citealt{Ueda2015}). This suggests that \srgas\ has spent most of the past several thousand years at similar levels of activity as found during our X-ray observations. 

Finally, we can estimate the luminosity of \srgas\ in the radio band (rest-frame 106--6980\,MHz) from the available measurements over the observed band of 74--4850\,MHz: $L_{\rm radio}\approx 2.3\times 10^{43}$\,erg\,s$^{-1}$, so that $L_{\rm radio}/\Lbol\approx 4\times 10^{-4}$. 

For comparison, for the \cite{shang2011} radio-loud and radio-quiet composite SEDs, which start at frequencies of 122\,MHz and 2100\,MHz respectively, we find $L_{\rm radio}/\Lbol\approx 4\times10^{-3} $ and $8\times10^{-7}$ (we estimated the corresponding bolometric luminosities as $5\lambda L_\lambda$\,(3000\AA), following \citealt{shang2011}). This again attests that \srgas\ is a radio-loud quasar. We finally note that giant radio lobes that are observed around FR II galaxies have estimated characteristic ages $\sim 100$\,Myr (e.g. \citealt{Machalski2009}), thus the presence of similar structures in \srgas\ indicates that substantial AGN activity in this object has maintained on at least a similar timescale.

%%%%%%%%%%%%%%%%%%%%%%%%%%%%%%%%%%%%%%%%%%%%%%%%
\section{Host galaxy and the central black hole}
\label{s:host}
%%%%%%%%%%%%%%%%%%%%%%%%%%%%%%%%%%%%%%%%%%%%%%%%

Since \srgas\ is a type 2 quasar, it is reasonable to assume that its SED in the optical and, perhaps, NIR bands is dominated by the emission of the host galaxy. This allows us to estimate its stellar mass. To this end we followed the methodologies outlined by \cite{Mendel2014}. Specifically, we modeled the SED using the Bayesian inference code Prospector \citep{Johnson2021}, which incorporates the Flexible Stellar Population Synthesis (FSPS) library \citep{Conroy2009} to generate synthetic stellar population models. We used the optical photometry from DESI LS DR 10, from which the contribution of emission lines (presumably emitted by the active nucleus) had been subtracted, and the NIR photometry from UHS DR2, all corrected for Galactic extinction, as was described in Section~\ref{s:opt_cont}.

\begin{figure}
\centering
\includegraphics[width=\columnwidth]{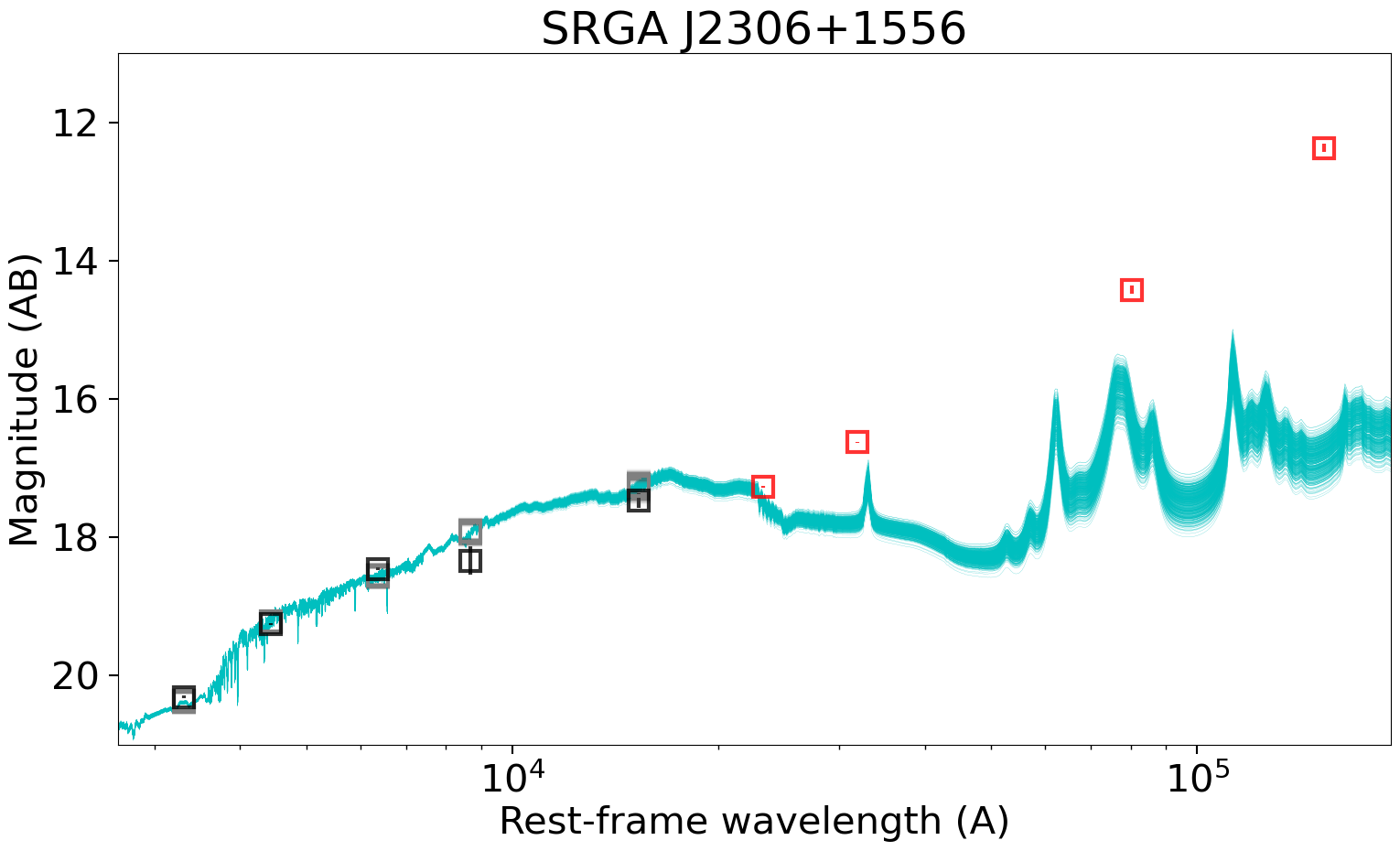}
\caption{
Modeling of the SED of \srgas's host galaxy. The black squares are the measured optical ($grz$) and NIR ($JK$) photometry corrected for Galactic extinction and with the contribution of optical emission lines having been subtracted. The gray squares show the best fit to these data, and the blue region shows 100 walkers of a Markov chain Monte Carlo analysis. The red squares are the MIR ($W1$, $W2$, $W3$, and  $W4$) WISE photometry, which was not used in the modeling, since it is presumably dominated by the emission from the AGN dusty torus.
}
\label{fig:sed_w}
\end{figure}

The best-fit parameters are summarized in Table~\ref{tab:host_properties}, where $\Mgal$ is the total stellar mass of the host galaxy, $u-r$ is the synthetic rest-frame color corrected for Galactic extinction, $\tausfh$ is the characteristic e-folding timescale of the star formation history, $\tage$ is the mean stellar age, $Z$ is the stellar metallicity, and $E(B-V)$ is the dust attenuation within the host galaxy. As can seen from Fig.~\ref{fig:sed_w}, the optical--NIR data are well described by this model, while the MIR part of the SED, not used in the analysis, rises dramatically above the extrapolation of the best-fit model. This excess emission can clearly be attributed to the AGN dusty torus. 
\begin{table}
\caption{
\label{tab:host_properties} 
Host galaxy properties inferred with SED fitting.
}
\setlength{\tabcolsep}{3pt}
\begin{tabular}{cccccc}
\toprule
\CellWithForceBreak{$\log\Mgal$, \\ $\Msun$ }
 & \CellWithForceBreak{$u-r$\\ mag} & \CellWithForceBreak{$\tausfh$\\ Gyr} & \CellWithForceBreak{$\tage$\\ Gyr}  & \CellWithForceBreak{$\log Z$\\ $Z_\odot$ } & \CellWithForceBreak{$E(B-V)$\\ mag}\\
\midrule
$11.39^{+0.03}_{-0.02}$ & $1.61^{+0.06}_{-0.06}$  & $5.78^{+1.16}_{-0.88}$  & $13.18^{+0.47}_{-1.08}$  & $0.48^{+0.02}_{-0.04}$  & $0.19^{+0.03}_{-0.07}$ \\

\bottomrule
\end{tabular} 
\begin{flushleft}
The uncertainties of the parameters are quoted at the $1\sigma$ level of confidence.
\end{flushleft}
\end{table}

We find that the stellar population is very old, with the mean star age, $13.2^{+0.5}_{-1.1}$\,Gyr, formally being larger than the age of the Universe (9.1\,Gyr) at $z=0.4389$. This probably indicates a more sophisticated history of star formation in the galaxy than is assumed in our (exponential) model. Also, perhaps, the NIR photometry is slightly affected by the AGN emission reprocessed by molecular gas and hot dust. Nevertheless, it is unlikely that our estimate of the total stellar mass is strongly affected by these systematic effects. We also note that the intrinsic reddening inferred for the host galaxy is significantly smaller than the value that we found for the NLR based on the Balmer decrement (see Sect.~\ref{s:opt_lines}), which suggests that the NLR is more dusty than the rest of the galaxy.

The inferred parameters of the galaxy, namely its large mass and old stellar population, suggest that it is an elliptical or at least bulge-dominated one. We can then estimate the mass of the central black hole, $\Mbh$, using the correlation between $\Mbh$ and the mass of the bulge, $\Mbulge$, in the local Universe \citep{Kormendy2013}:
\begin{equation}
\frac{\Mbh}{10^{9}\Msun}=\left(0.49^{+0.06}_{-0.05}\right)\left(\frac{\Mbulge}{10^{11}\Msun}\right)^{1.17\pm 0.08}, 
\label{eq:mbh}
\end{equation}
Substituting the $\Mgal$ value from Table~\ref{tab:host_properties} for $\Mbulge$ in this formula, we find $\Mbh\sim 1.4\times 10^9$\,$\Msun$ with a factor of $\sim 2$ uncertainty, mainly due to the intrinsic scatter in the $\Mbh$--$\Mbulge$ correlation (0.28\,dex, \citealt{Kormendy2013}). 

Finally, using our estimate of the bolometric luminosity of $\Lbol\sim (6\pm 1)\times 10^{46}$\,erg\,s$^{-1}$, we find that it constitutes $\sim 30$\% of the object's Eddington luminosity for the derived black hole mass. Therefore, the quasar accretes mass at a high rate.

%%%%%%%%%%%%%%%%%%%%%%%%%%%%%%%%%%%%%%%%%%%%%%%%%%%%%%%%%%%%%%%
\section{\srgas\ in the AGN population context and conclusions}
\label{s:population}
%%%%%%%%%%%%%%%%%%%%%%%%%%%%%%%%%%%%%%%%%%%%%%%%%%%%%%%%%%%%%%%

\begin{figure}
\centering
\includegraphics[width=\columnwidth]{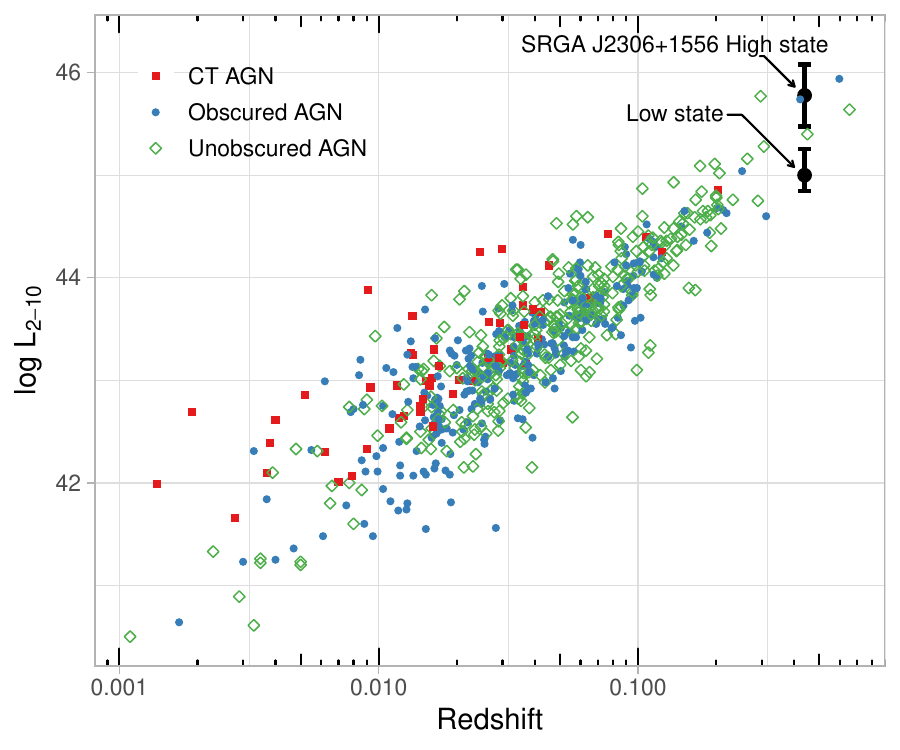}
\caption{
Intrinsic 2--10\,keV luminosity vs. redshift for the AGN from the 70-month \swift/BAT catalog, excluding blazars. The sources are classified as unobscured ($\NH < 10^{22}$cm$^{-2}$), obscured ($10^{22}$ < $\NH$ < $10^{24}$ cm$^{-2}$), and Compton-thick ($\NH>10^{24}$\,cm$^{-2}$). The corresponding uncertainties are not shown. For comparison, we show \srgas\ in its low and high luminosity states, respectively, and the corresponding uncertainties (see the legend). 
}
\label{fig:lx_z}
\end{figure}

\srgas\ proves to be a fairly exceptional AGN for the $z<0.5$ Universe. First, its unabsorbed X-ray luminosity ranged between $\Lx=1.0^{+0.8}_{-0.3}\times 10^{45}$ and $6^{+6}_{-3}\times 10^{45}$\,erg\,s$^{-1}$ (2--10\,keV) in our observations, and it exhibited a strongly absorbed X-ray spectrum, with $\NH\sim 2\times 10^{23}$\,cm$^{-2}$. Such X-ray luminous obscured AGN are extremely rare in the low-redshift Universe. 

We can best judge about this based on the serendipitous all-sky surveys that have been conducted over the past two decades in hard X-rays (at energies above 15\,keV), where absorption bias is minimal, by the BAT instrument aboard the \swift\ observatory (e.g. \citealt{Oh2018}) and the IBIS instrument aboard {\it INTEGRAL} (e.g. \citealt{Krivonos2022}). Thanks to follow-up X-ray and optical identification programs, these surveys have provided highly complete catalogs of hard X-ray selected AGN with information on their redshifts, optical types, and X-ray spectral properties. Figure \ref{fig:lx_z} shows the location of objects from the \swift/BAT 70-month AGN catalog \citep{Ricci2017} on a diagram of redshift vs. intrinsic luminosity in the 2--10\,keV band. Here, the blazars have been excluded and a distinction is made between unobscured ($\NH<10^{22}$\,cm$^{-2}$), obscured ($10^{22}$ < $\NH$ < $10^{24}$ cm$^{-2}$), and Compton-thick ($\NH>10^{24}$\,cm$^{-2}$) AGN. The intrinsic luminosities in this catalog were determined from broad-band X-ray spectral analysis of data from various X-ray telescopes, using the same cosmology as ours and more accurate distances for the nearest AGN. For comparison, we placed \srgas\ on the same diagram, using its redshift $z=0.4389$ and our luminosity estimates for the low and high states.

There are just 11 non-blazar AGN with $\Lx > 10^{45}$\,erg\,s$^{-1}$ in the \swift/BAT catalog. They are located at $z=0.1735$--0.6540 and only three of them are obscured: Swift\,J0952.3$-$6234 ($z=0.252$, $\Lx\approx 1.1\times 10^{45}$\,erg\,s$^{-1}$, $\NH\approx 1.2\times 10^{23}$\,cm$^{-2}$), Swift\,J0216.3+5128 ($z=0.422$, $\Lx\approx 5.5\times 10^{45}$\,erg\,s$^{-1}$, $\NH\approx 2.8\times 10^{22}$\,cm$^{-2}$), and Swift\,J2344.6$-$4246 ($z=0.5975$, $\Lx\approx 8.7\times 10^{45}$\,erg\,s$^{-1}$, $\NH\approx 3.3\times 10^{23}$\,cm$^{-2}$). 

Of course, the \swift/BAT all-sky survey is a fairly shallow one, with a sensitivity of $1.3\times 10^{-11}$\,erg\,s$^{-1}$ or better in the 14--195\,keV energy band over 90\% of the sky for the 70-month survey \citep{Baumgartner2013}. This, assuming a power-law spectrum with $\Gamma=1.8$ and a $E_{\rm cut}=150$\,keV, implies that 
quasars with $\Lx>3\times 10^{45}$\,erg\,s$^{-1}$ (such as \srgas\ in its high state) are robustly detectable by \swift/BAT out to $z=0.336$. Therefore, the presence of just one such high-luminosity AGN up to this redshift in the \swift/BAT catalog (Swift\,J1822.0+6421, $z=0.297$, $\Lx\approx5.9\times 10^{45}$\,erg\,s$^{-1}$, unobscured) 
means that it is indeed alone in the entire low-redshift Universe.

Apart from being X-ray luminous, \srgas\ is also radio-loud. 
We can classify it as a giant FRII radio galaxy based on its aforementioned morphological features and radio power. \srgas\ well fits into the theoretical framework of evolution of double extragalactic radio sources (e.g. \citealt{An2012}). We observe this source in the stage of Large Symmetric Objects (LSO) nowadays, during which the radio spectrum is steep with an index $\alpha\sim-1$. Inverse Compton losses resulting from the interaction with the Cosmic microwave background presumably prevail over synchrotron losses at this stage. 

In fact, apart from X-ray surveys, luminous obscured quasars can also be discovered in low-frequency radio surveys, which are virtually unaffected by intrinsic absorption. A large sample of FR II galaxies selected from the 3CRR catalog \citep{Laing1983} up to $z\sim2.5$ has been observed with modern soft X-ray telescopes \citep[see, e.g.][]{Massaro2015, Jimenez-Gallardo2020}. Utilizing data from this sample \cite{Kuraszkiewicz2021} have shown that luminous narrow-line radio galaxies (NLRGs), resembling \srgas, are fairly frequent at $0.5 < z < 1$. This is consistent with AGN statistics inferred from deep (pencil-beam) X-ray surveys, which probe the high-redshift Universe. The measured AGN X-ray luminosity function (XLF) shows that the number density of luminous ($\Lx\gtrsim 10^{45}$\,erg\,s$^{-1}$) quasars increases rapidly with redshift up to $z\sim 2$ and that $\sim 20$\% (this fraction tends to decrease with increasing luminosity) of them are obscured (e.g. \citealt{Ueda2014}).

Therefore, our discovery of a powerful obscured quasar (or, equivalently, a NLRG) at $z\approx 0.44$ has evidently become possible due to a combination of two factors: i) strong cosmological evolution of the XLF and ii) higher sensitivity (for moderately obscured AGN, $\NH\lesssim 2\times 10^{23}$\,cm$^{-2}$) of the \srg/\art\ all-sky survey compared to the \swift/BAT and {\it INTEGRAL}/IBIS all-sky surveys. The latter allowed us to reach further out into the Universe and thus probe a significantly larger comoving volume, which resulted in detecting one of only few luminous obscured quasars that have existed over the last 5 billion years.

In conclusion, \srgas\ is one of the nearest luminous obscured quasars, similar in its intrinsic properties to powerful quasars and NLRGs that become abundant at significantly higher redshifts. As such, it can serve as a valuable testbed for in-depth exploration of the physics of such objects. Further studies of \srgas\ could include its broader band X-ray spectroscopy to better constrain the spectral shape (in particular, to measure a high energy cutoff due to Compton recoil), low-frequency radio mapping to better expose the large-scale structures, and continued monitoring in X-ray, optical, IR, and radio bands to reveal luminosity and spectral variability.

\section*{Acknowledgements}

This work is based on observations with the Mikhail Pavlinsky ART-XC telescope, the hard X-ray instrument on board the \srg\ observatory. The \srg\ observatory was created by Roskosmos in the interests of the Russian Academy of Sciences represented by its Space Research Institute (IKI) in the framework of the Russian Federal Space Program, with the participation of
Germany. The ART-XC team thanks Lavochkin Association (NPOL) with partners for the creation and operation of the \srg\ spacecraft (Navigator). 

This work also made use of data supplied by the UK Swift Science Data Centre at the University of Leicester; data of the Sloan Digital Sky Survey IV, provided by the Alfred P. Sloan Foundation, the U.S. Department of Energy Office of Science, and the Participating Institutions; data obtained with the Dark Energy Spectroscopic Instrument (DESI), whose construction and operations is managed by the Lawrence Berkeley National Laboratory; and data of the Very Large Array Sky Survey (VLASS) provided by the National Radio Astronomy Observatory and the Canadian Initiative for Radio Astronomy Data Analysis. 

This research was supported by the Ministry of Science and Higher Education grant 075-15-2024-647.

%%%%%%%%%%%%%%%%%%%%%%%%%%%%%%%%%%%%%%%%%%%%%%%%%%

\section*{Data Availability}
The \swift/XRT, SDSS, DESI, and VLASS data used in this article are accessible through the corresponding web pages.
At the time of writing, the \srg/\art\ data and the corresponding data analysis software have a private status. We plan to provide public access to the \art\ scientific archive in the future.

%%%%%%%%%%%%%%%%%%%% REFERENCES %%%%%%%%%%%%%%%%%%

% The best way to enter references is to use BibTeX:

\bibliographystyle{mnras}
\bibliography{bibl} % if your bibtex file is called example.bib

% Alternatively you could enter them by hand, like this:
% This method is tedious and prone to error if you have lots of references
%\begin{thebibliography}{99}
%\bibitem[\protect\citeauthoryear{Author}{2012}]{Author2012}
%Author A.~N., 2013, Journal of Improbable Astronomy, 1, 1
%\bibitem[\protect\citeauthoryear{Others}{2013}]{Others2013}
%Others S., 2012, Journal of Interesting Stuff, 17, 198
%\end{thebibliography}

%%%%%%%%%%%%%%%%%%%%%%%%%%%%%%%%%%%%%%%%%%%%%%%%%%

%%%%%%%%%%%%%%%%% APPENDICES %%%%%%%%%%%%%%%%%%%%%

%\appendix
%\section{Some extra material}

%%%%%%%%%%%%%%%%%%%%%%%%%%%%%%%%%%%%%%%%%%%%%%%%%%

% Don't change these lines
\bsp	% typesetting comment
\label{lastpage}
\end{document}